\documentclass[12pt]{article}
\usepackage{tipa}
\usepackage[all,import]{xy}
\usepackage{graphicx,color}
\usepackage{amsmath}
\usepackage{amsbsy}
\usepackage{amssymb}
\usepackage{cases}
\usepackage{enumerate}
\usepackage{bm}
\usepackage[colorlinks,linkcolor=black,anchorcolor= black,citecolor= black,urlcolor=black]{hyperref}
\usepackage{amsthm}
\usepackage{multirow}
\usepackage{subfigure}
\usepackage{colortbl}
\usepackage{kbordermatrix}
\usepackage{algorithm}
\usepackage{algorithmic}

\definecolor{mygray}{gray}{.9}

\makeatletter
\newcommand*{\rom}[1]{\expandafter\@slowromancap\romannumeral #1@}
\makeatother

\begin{document}

\theoremstyle{definition}
\newtheorem{assumption}{Assumption}
\newtheorem{theorem}{Theorem}
\newtheorem{lemma}{Lemma}
\newtheorem{example}{Example}
\newtheorem{definition}{Definition}
\newtheorem{corollary}{Corollary}

\def\letas{\mathrel{\mathop{=}\limits^{\triangle}}}
\def\ind{\begin{picture}(9,8)
         \put(0,0){\line(1,0){9}}
         \put(3,0){\line(0,1){8}}
         \put(6,0){\line(0,1){8}}
         \end{picture}
        }
\def\nind{\begin{picture}(9,8)
         \put(0,0){\line(1,0){9}}
         \put(3,0){\line(0,1){8}}
         \put(6,0){\line(0,1){8}}
         \put(1,0){{\it /}}
         \end{picture}
    }

\def\var{\text{var}}
\def\cov{\text{cov}}
\def\sumn{\sum\limits_{i=1}^n}
\def\summ{\sum\limits_{j=1}^m}
\def\convergeas{\stackrel{a.s.}{\longrightarrow}}
\def\converged{\stackrel{d}{\longrightarrow}}
\def\iidsim{\stackrel{i.i.d.}{\sim}}
\def\indsim{\stackrel{ind}{\sim}}
\def\asim{\stackrel{a}{\sim}}
\def\d{\text{d}}

\setlength{\baselineskip}{1.5\baselineskip}

\title{\bf Bayesian Robust Inference of Sample Selection Using Selection-t Models}
\author{Peng Ding \\
Department of Statistics, Harvard University\\
Address: One Oxford Street, Cambridge, MA 02138, USA\\
E-mail: \href{mailto:pengding@fas.harvard.edu}{pengding@fas.harvard.edu}\\
 }
\date{}
\maketitle


\begin{center}
{\bf Abstract}
\end{center}
\quad Heckman selection model is the most popular econometric model in analysis of data with sample selection.
However, selection models with Normal errors cannot accommodate heavy tails in the error distribution.
Recently, Marchenko and Genton proposed a selection-t model to perform frequentist' robust analysis of sample selection.
Instead of using their maximum likelihood estimates, our paper develops new Bayesian procedures for the selection-t models with either continuous or binary outcomes. 
By exploiting the Normal mixture representation of the t distribution, we can use
data augmentation to impute the missing data,
and use parameter expansion to sample the restricted covariance matrices. 
The Bayesian procedures only involve simple steps, 
without calculating analytical or numerical derivatives of the complicated log likelihood functions. 
Simulation studies show the vulnerability of the selection models with Normal errors, as well as the robustness of the selection models with t errors. 
Interestingly, we find evidence of heavy-tailedness in three real examples analyzed by previous studies, and the conclusions about the existence of selection effect are very sensitive to the distributional assumptions of the error terms.


\newpage

\section{Introduction}

Sample selection often occurs in social sciences and biomedical studies, when the outcomes of interest are partially observed or the samples are not representative of the population.
In the analysis of labor market, Heckman (1979) proposed a selection model comprising of a Probit sample selection equation and a Normal linear outcome equation. 
The sample selection problem arises, when the error terms of the sample selection equation and the outcome equation are correlated.
Heckman (1979) treated the sample selection as a model misspecification problem due to an omitted variable, and 
proposed a two-step procedure to adjust the linear regression model with an extra nonlinear term.
The Heckman selection model, or Type \rom{2} Tobit model, is now widely used in many fields.
Heckman's two-step procedure and the maximum likelihood estimation (MLE) using Newton-Raphson iteration are incorporated into many standard econometrical and statistical routines, such as the \texttt{heckman} procedure in \texttt{Stata} (StataCorp 2013) and the \texttt{sampleSelection} (Toomet and Henningsen 2008) package of \texttt{R} (R Development Core Team 2010).

Despite its popularity and wide applications, researchers tried to generalize the Heckman selection model in various ways. Examples include
transformation-based model (Lee 1983), semiparametric model (Ahn and Powell 1993), and nonparametric model (Das, Newey and Vella 2003).
Among the Bayesian community, 
Li (1998) proposed a Bayesian inference procedure for the Heckman selection model using data augmentation (Tanner and Wong 1987), Chib, Greenberg and Jeliazkov (2009) generalized it to
Bayesian nonparametric model, and Van Hasselt (2011) generalized it to Bayesian semiparametric model using Dirichlet process prior.

Models based on t distributions are widely applied for robust analysis (Albert and Chib 1993; Geweke 1992; Liu 1999; Liu 2004),
and they are attractive alternatives for the models based on Normal distribution such as linear and Probit models.
Recently, Marchenko and Genton (2012) extended
the selection model to deal with heavy-tailedness by modeling the error terms as a bivariate t distribution, which was called a selection-t model.
Marchenko and Genton (2012) proposed a Newton-Raphson iteration procedure to find the MLE of the selection-t model. However, the frequentists' solution has several limitations: first, it involves complicated derivatives of the log likelihood function; second, it is not very direct to be generalized to binary outcomes; third, the inference based on the asymptotic Normality of MLE 
may not be accurate in problems with small sample sizes.
In order to overcome these limitations, we propose
Bayesian procedures for the selection-t model using Markov Chain Monte Carlo (MCMC). The Bayesian procedures exploit the Normal mixture representation of the t distribution (Albert and Chib 1993), and use data augmentation (Tanner and Wong 1987) to impute the latent variables.
However, difficulty arises when sampling the covariance matrix of the error terms.
For the purpose of full identification, the variance of the error term in the selection equation is restricted to be one, which makes the posterior distribution of the covariance matrix non-standard and difficult to sample directly. Previous studies (Koop and Poirier 1997; Li, 1998; McCulloch, Polson and Rossi 2000; Van Hasselt 2011) reparametrized the restricted covariance matrix.
However, we use parameter expansion (Liu and Rubin 1998; Meng and Van Dyk 1999; Van Dyk and Meng 2001;  Imai and Van Dyk 2005) to overcome the sampling difficulty of the restricted covariance matrix, which allows us to have conjugate distributions. 
We provide a more detailed discussion of the two classes of approaches in Section \ref{sec::discussion}.

The remainder of the article proceeds as follows. 
In Section \ref{sec::heckman}, we briefly review the Heckman selection model. 
Section \ref{sec::selectiont} introduces the selection-t model (Marchenko and Genton 2012) and its Normal mixture representation, which is fundamental to our Bayesian procedures.
Section \ref{sec::bayesian} presents a Bayesian inference procedure for the selection-t model, and
Section \ref{sec::robit} generalizes it to deal with binary outcome.
We show some simulation studies to evaluate the finite sample properties of our procedures in Section \ref{sec::simulation}.
In Section \ref{sec::applications}, we apply our new Bayesian procedures to empirical applications, where we find
evidence of heavy-tailedness.
Section \ref{sec::discussion} contains some discussions and possible extensions.
All the technical details are shown in Appendix.
Throughout this article, all vectors are column vectors, and we use boldface letters to represent vectors and matrices. 
Functions written in \texttt{R} for the methods proposed in this paper are available upon request from the author.

\section{Review of the Heckman Selection Model}\label{sec::heckman}
A selection model has two parts: a regression equation for the outcome, and a regression equation for the sample selection mechanism.
Suppose the regression equation for the outcome of primary interest is
\begin{eqnarray}\label{eq::outcome}
y_i^* = \bm{x}_i^\top \bm{\beta} + \varepsilon_i,
\end{eqnarray}
and the sample selection mechanism is driven by the following latent linear regression equation
\begin{eqnarray}\label{eq::selection}
u_i^* = \bm{w}_i^\top \bm{\gamma} + \eta_i,
\end{eqnarray}
for $i = 1, \cdots, N$.
The covariates in $\bm{x}_i$ and $\bm{w}_i$ may overlap with each other, and the exclusion restriction holds when at least one of the elements of $\bm{w}_i$ are not in $\bm{x}_i$.
Let $K$ and $L$ denote the dimensions of $\bm{x}_i$ and $\bm{w}_i$, respectively.
We observe the outcome $y_i^*$, if and only if $u_i^* > 0$. Therefore, the indicator for sample selection is
\begin{eqnarray}
u_i = I(u_i^* > 0 ).
\end{eqnarray}
Let $y_i$ be the observed outcome, with $y_i = y_i^*$ if $u_i = 1$, and $y_i = \text{NA}$ is $u_i = 0$, where ``NA'' indicates missing data.

Heckman (1979) assumed a bivariate Normal distribution for $\varepsilon_i$ and $\eta_i$:
\begin{eqnarray}\label{eq::n}
\begin{pmatrix}
\varepsilon_i\\
\eta_i
\end{pmatrix}
\sim
\bm{N}_2( 
\bm{0}_2,
\bm{\Omega}
),
\end{eqnarray}
where $\bm{0}_2 = \begin{pmatrix}  0\\0 \end{pmatrix} $ and $\bm{\Omega} =\begin{pmatrix}   \sigma^2_1 & \rho\sigma_1 \\ \rho\sigma_1 & 1\end{pmatrix} $. We fix the second diagonal element of $\bm{\Omega}$ at $1$ for full identification.
Under the bivariate Normal assumption, the mean equation for the outcomes of the selected samples is
\begin{eqnarray}\label{eq::mills}
E\{y\mid u =1, \bm{x}, \bm{w} \} = \bm{x}^\top \bm{\beta} + \rho\sigma_1 \lambda(\bm{w}^\top \bm{\gamma}),
\end{eqnarray}
where $\lambda(\cdot) = \phi(\cdot)/\Phi(\cdot)$ is the inverse Mills ratio.
Therefore, the sample selection problem can be treated as a model misspecification problem,
because the mean equation for the outcomes of the selected samples is a linear function $\bm{x}_i^\top\bm{\beta}$ with a nonlinear correction term $\rho\sigma_1 \lambda(\bm{w}^\top \bm{\gamma})$.
Based on (\ref{eq::mills}), Heckman (1979) proposed a two-step procedure by first fitting a Probit model of $u$ on $\bm{w}$ to obtain $\widehat{\bm{\gamma}}$, and then fitting a linear model of $y$ on $\left\{  \bm{x}, \lambda(\bm{w}^\top \widehat{ \bm{\gamma}  } )  \right\}$ to obtain $( \widehat{\bm{\beta}},  \widehat{\rho}, \widehat{\sigma}_1)$.
The two-step procedure is less efficient than the full information MLE, but it is robust to the deviation of the joint Normality of the error terms.
The MLE of the Heckman selection model can be calculated by Newton-Raphson iteration or EM algorithm (Little and Rubin 2002). 
Alternatively, Bayesian posterior inference of the Heckman selection model can be obtained by data augmentation (Li 1998).


\section{Normal Mixture Representation of the Selection-t Model}\label{sec::selectiont}

In order to model heavy-tailedness, Marchenko and Genton (2012) proposed a selection-t model, and assumed that $\varepsilon_i$ and $\eta_i$ follow a bivariate t distribution with unknown degrees of freedom $\nu$, namely,
\begin{eqnarray}\label{eq::t}
\begin{pmatrix}
\varepsilon_i\\
\eta_i
\end{pmatrix}
\sim \bm{t}_2 (
\bm{0}_2, \bm{\Omega},
\nu
).
\end{eqnarray}
The density function of the bivariate t distribution $\bm{t}_2( \bm{\mu}, \bm{\Omega},  \nu)$ is
\begin{eqnarray}
f(\bm{t}; \bm{\mu}, \bm{\Omega},  \nu) = (2\pi)^{-1} |\bm{\Omega}|^{-1/2} 
\left\{   1 +\nu^{-1} (   \bm{t} - \bm{\mu}     )^\top \bm{\Omega}^{-1} (   \bm{t} - \bm{\mu}     )  \right\}^{-(\nu + 2)/2}.\label{eq::t_pdf}
\end{eqnarray}
As $\nu\rightarrow +\infty$, the bivariate t distribution in (\ref{eq::t}) converges to the bivariate Normal distribution in (\ref{eq::n}). Thus, the Heckman selection model is a limiting case of the selection-t model.
In this article, we use the name ``selection model'' for (\ref{eq::n}) and the name ``selection-t model'' for (\ref{eq::t}). 

However,
the density of the t distribution in (\ref{eq::t_pdf}) results in cumbersome posterior distributions, which can be solved by using data augmentation.
By introducing latent variables $\{q_i:i=1, \cdots, N\}$,
the bivariate t distribution of $\varepsilon_i$ and $\eta_i$ has the following Normal mixture representation:
\begin{eqnarray}\label{eq::mix}
\begin{pmatrix}
\varepsilon_i\\
\eta_i
\end{pmatrix}
\sim  \bm{ N}_2(\bm{0}_2, 
\alpha 
\bm{\Omega}/ q_i ), \text{ where }q_i\sim \alpha \chi^2_{\nu}/ \nu, i = 1, \cdots, N.
\end{eqnarray}
The parameter $\alpha$ is not identifiable from the observed data $\bm{D}_{obs}= \{ (y_i, u_i, \bm{x}_i, \bm{w}_i):i=1, \cdots, N\}$, but it is identifiable from the complete data $\bm{D}_{com} = \{  (y_i^*, u_i^*, y_i, u_i, q_i, \bm{x}_i, \bm{w}_i): i = 1, \cdots, N  \}$.
When $\alpha$ is fixed at one, the model is fully identifiable.
The overparametrization for this model is a way of parameter expansion to accelerate the convergence rates of the MCMC samplers
(Liu and Wu 1999; Meng and Van Dyk 1999; Van Dyk and Meng 2001).

For Bayesian inference, we need to specify prior distributions for all the parameters $ (\bm{\beta}, \bm{\gamma}, \bm{\Omega}, \nu, \alpha)$.
We choose a multivariate Normal prior for the regression coefficients
$
(\bm{\beta} ^\top, \bm{\gamma}^\top)^\top  \sim \bm{N}_{K+L}( \bm{\mu}_0, \bm{\Sigma}_0 ),
$
a Gamma prior for the degrees of freedom, $\nu\sim \text{Gamma} (\alpha_0, \beta_0)$ with a shape parameter $\alpha_0$ and a rate parameter $\beta_0$, and
a scaled-inverse-$\chi^2$ prior for $\alpha$, $\alpha\sim b/\chi^2_{c}$.

We restrict the second diagonal element of the covariance matrix $\bm{\Omega}$ to be one, 
which makes it difficult to sample $\bm{\Omega}$ from its posterior distribution directly. 
In order to use parameter expansion, we consider the unrestricted covariance matrix 
$\bm{\Sigma} = \text{diag}\{1, \sigma_2\} ~ \bm{\Omega} ~\text{diag}\{1, \sigma_2\}$. The Inverse-Wishart prior $\bm{W}_2^{-1}(\nu_0, \bm{I}_2)$ for the covariance matrix $\bm{\Sigma}$ is equivalent to the priors for $( \bm{\Omega}, \sigma_2^2)$:
\begin{eqnarray}
f(\bm{\Omega}) &\propto&  (1 - \rho^2)^{-3/2}  \sigma_1^{-(\nu_0 + 3)} \exp\left\{   -\frac{1}{ 2\sigma^2_1 (1 - \rho^2) }   \right\},  \label{eq::prior_omega} \\
\text{ and }\sigma_2^2|\bm{\Omega}&\sim& \{  (1 - \rho^2) \chi^2_{\nu_0}  \}^{-1}.  \label{eq::prior_sigma2}
\end{eqnarray}
Details of the derivations for (\ref{eq::prior_omega}) and (\ref{eq::prior_sigma2}) are in Appendix A.

\section{Bayesian Computation for the Selection-t Model}\label{sec::bayesian}

Bayesian computation using data augmentation includes two main steps: the imputation step (I-step) by imputing the missing data $\bm{D}_{mis} = \{  (y_i^*,u_i^*, q_i ):i=1, \cdots, N \}$, and the posterior step (P-step) by updating the posterior distributions of the parameters.

For the ease of derivation, we introduce the following matrix notation.
The joint model of the latent outcome and the selection mechanism is
\begin{eqnarray*}
\begin{pmatrix}  y_i^* \\ u_i^* \end{pmatrix}  
= \begin{pmatrix}  \bm{x}_i^\top & \bm{0}_L \\ \bm{0}_K & \bm{w}_i^\top \end{pmatrix} 
\begin{pmatrix}   \bm{\beta} \\ \bm{\gamma}  \end{pmatrix}
+   \begin{pmatrix} \varepsilon_i \\ \eta_i \end{pmatrix} .
\end{eqnarray*}

Define $ \bm{Z}_i =  \begin{pmatrix}  y_i^* \\ u_i^* \end{pmatrix} $ as the latent outcome and selection mechanism, $ \bm{V}_i  =  \begin{pmatrix}  \bm{x}_i^\top & \bm{0}_L \\ \bm{0}_K & \bm{w}_i^\top \end{pmatrix}  $ as the design matrix of the covariates, and $ \bm{\delta} = \begin{pmatrix}     \bm{\beta} \\ \bm{\gamma} \end{pmatrix}$ as the regression coefficients.
The complete data likelihood is
\begin{eqnarray}
&&\alpha ^{ - n} |\bm{\Omega}|^{-n/2} 
 \exp\left\{    - \frac{1}{2\alpha} \sum_{i=1}^N q_i (\bm{Z}_i - \bm{V}_i \bm{\delta})^\top \bm{\Omega}^{-1}  (\bm{Z}_i - \bm{V}_i \bm{\delta})     \right\}  \nonumber \\
 &&\cdot \prod_{i=1}^n  q_i \frac{1}{ (2\alpha/\nu)^{\nu/2} \Gamma(\nu/2)  }  q_i^{\nu/2-1}   e^{-\nu q_i / (2\alpha) }  \nonumber \\
&\propto&   \alpha^{-n - n\nu/2}  (\nu/2)^{ n\nu/2} \{ \Gamma(\nu/2) \}^{-n}  |\bm{\Omega}|^{-n/2} 
 \exp\left\{    - \frac{1}{2\alpha} 
 \sum_{i=1}^N q_i (\bm{Z}_i - \bm{V}_i \bm{\delta})^\top \bm{\Omega}^{-1}  (\bm{Z}_i - \bm{V}_i \bm{\delta})     \right\}  \nonumber\\
 &&\cdot \prod_{i=1}^n  q_i^{\nu/2}   e^{-\nu q_i / (2\alpha) }. \label{eq::lcom}
\end{eqnarray}

\subsection{The Imputation Step}
First, we impute the missing data given the observed data and the parameters.
Let $TN(\mu, \sigma^2; L, U)$ be a Normal distribution 
$N(\mu, \sigma^2)$ truncated within the interval $[L, U]$. 
In the imputation step, we sample $ \left(\alpha, \{\bm{Z}_i , q_i\} | \{y_i, u_i \} , \bm{\delta}, \bm{\Omega}, \nu\right)$ jointly, and then marginalize over $\alpha$ by discarding its sample.
Since $\alpha |(  \{y_i, u_i \} , \bm{\delta}, \bm{\Omega} , \nu ) $ is the same as its prior distribution, we draw $\alpha\sim b/\chi^2_c$.
Given $( u_i^* , y_i, u_i , \bm{\delta}, \bm{\Omega}, \nu, \alpha )$, we use Gibbs sampler to draw $\{ \bm{Z}_i\}$ and $ \{ q_i\} $ iteratively. 
The conditional means and variances of the bivariate Normal variables $( y_i^* , u_i^*  )$ have the following forms:
\begin{eqnarray}
&\mu_{u|y} = \bm{w}_i^\top \bm{\gamma}  +  \rho    (y_i^* - \bm{x}_i^\top \bm{\beta}) / \sigma_1,
&\sigma^2_{u|y} = \alpha(1 - \rho^2)/q_i,\\
&\mu_{y|u} = \bm{x}_i^\top \bm{\beta} + \rho\sigma_1 (u_i^* - \bm{w}_i^\top \bm{\gamma}  ),
&\sigma^2_{y|u} = \alpha\sigma_1^2(1 - \rho^2)/q_i.
\end{eqnarray}
Given $(q_i, y_i, u_i , \bm{\delta}, \bm{\Omega}, \nu, \alpha)$, we impute the latent variables $(y_i^*, u_i^*)$ as follows:
if $u_i = 1$, we draw 
\begin{eqnarray}\label{eq::mis1}
&&y_i^* = y_i,\\
 \text{ and }&&
u_i^*| \left(  y_i^* , q_i, y_i, u_i , \bm{\delta}, \bm{\Omega}, \nu, \alpha \right)  \sim TN( \mu_{u|y}, \sigma^2_{u|y}; 0, \infty ) ;
\end{eqnarray}
if $u_i = 0$, we draw 
\begin{eqnarray}
&&u_i^*| \left(  q_i,  y_i, u_i , \bm{\delta}, \bm{\Omega},\nu,  \alpha  \right)  \sim TN( \bm{w}_i^\top \bm{\gamma}, \alpha/q_i; -\infty, 0 ),\\
\text{ and }&&
y_i^*| \left(  u_i^* , q_i , y_i, u_i , \bm{\delta}, \bm{\Omega}, \nu, \alpha \right) \sim N( \mu_{y|u}, \sigma_{y|u}^2  )  .\label{eq::mis0}
\end{eqnarray}

Given $(\bm{Z}_i, y_i, u_i  , \bm{\delta}, \bm{\Omega}, \nu, \alpha)$, we impute $q_i$ by the scaled-inverse-$\chi^2$ distribution:
\begin{eqnarray}\label{eq::q}
q_i    | \left(   \bm{Z}_i, y_i, u_i  , \bm{\delta}, \bm{\Omega}, \nu, \alpha \right)  \sim 
{  \alpha \chi^2_{\nu+2}   / \left\{ (\bm{Z}_i - \bm{V}_i \bm{\delta})^\top \bm{\Omega}^{-1}  (\bm{Z}_i - \bm{V}_i \bm{\delta}) + \nu     \right\} }.
\end{eqnarray}

Algorithm \ref{alg::impute_t} summarizes the imputation step for the selection-t model.
\begin{algorithm}
\caption{Imputation step for the selection-t model}
\label{alg::impute_t}
\begin{algorithmic}
\item[I-1] Draw $\alpha \sim b/\chi^2_c$ from its prior distribution;

\item[I-2] Draw $(y_i^*, u_i^*)$ according to (\ref{eq::mis1}) to (\ref{eq::mis0}) for $i= 1, \cdots, N$;

\item[I-3] Draw $ q_i   $ according to (\ref{eq::q}) for $i=1, \cdots, N$.
\end{algorithmic}
\end{algorithm}

\subsection{Posterior Step}
Second, we draw the parameters from their posterior distributions, conditioning on the complete data $\bm{D}_{com}$.
After imputing the missing data, the parameter $\alpha$ is identifiable from the complete data, and it follows a scaled-inverse-$\chi^2$ distribution:
\begin{eqnarray}\label{eq::alpha}
\alpha | \left(  \{  \bm{Z}_i,q_i,  \bm{V}_i \}, \bm{\delta}, \bm{\Omega}, \nu \right)  \sim  \left[   b + \sum_{i=1}^N q_i \{  (\bm{Z}_i - \bm{V}_i \bm{\delta})^\top \bm{\Omega}^{-1}  (\bm{Z}_i - \bm{V}_i \bm{\delta})     + \nu  \}    \right]\Big/\chi^2_{c + 2N + N\nu}  .
\end{eqnarray}

The complete data likelihood in (\ref{eq::lcom}) demonstrates the Normality of $\bm{\delta} | \left( \{  \bm{Z}_i, \bm{V}_i\} , \bm{\Omega}, \nu, \alpha  \right)$, because of the quadratic log posterior density.
The posterior mean and precision matrix of $\bm{\delta}$ are determined by the mode and the negative Hessian matrix of the log posterior density.
Since the log of the conditional posterior density of $\bm{\delta}$ is
\begin{eqnarray*}
 - \frac{1}{2\alpha} \sum_{i=1}^N q_i (\bm{Z}_i - \bm{V}_i \bm{\delta})^\top \bm{\Omega}^{-1}  (\bm{Z}_i - \bm{V}_i \bm{\delta})   - \frac{1}{2} (\bm{\delta} - \bm{\mu}_0 )^\top \bm{\Sigma}_0^{-1} ( \bm{\delta} - \bm{\mu}_0),
\end{eqnarray*}
we draw 
\begin{eqnarray}\label{eq::delta}
 \bm{\delta} | \left(  \{  \bm{Z}_i, \bm{V}_i\} , \bm{\Omega}, \nu, \alpha  \right) \sim \bm{N}_{K+L}( \widehat{\bm{\mu}}_{\bm{\delta}},   \widehat{\bm{\Sigma}}_{\bm{\delta}} ),
 \end{eqnarray}
where
\begin{eqnarray*}
\widehat{\bm{\mu}}_{\bm{\delta}} =  \widehat{\bm{\Sigma}}_{\bm{\delta}} \left(    \sum_{i=1}^N q_i \bm{V}_i^\top \bm{\Omega}^{-1}\bm{Z}_i/\alpha + \bm{\Sigma}_0^{-1}\bm{\mu}_0      \right)
\text{ and }
\widehat{\bm{\Sigma}}_{\bm{\delta}} =  \left(    \sum_{i=1}^N  q_i \bm{V}_i^\top \bm{\Omega}^{-1}\bm{V}_i/\alpha + \bm{\Sigma}_0^{-1}   \right)^{-1}.
\end{eqnarray*}

Given $(\{\bm{Z}_i, \bm{V}_i,  q_i\},  \bm{\delta} ,  \nu, \alpha  )$, it is difficult to sample the restricted covariance matrix $\bm{\Omega}$ directly.
However, parameter expansion allows us to reparametrize the model and get conjugate posterior distributions.
Define
\begin{eqnarray}\label{eq::trans}
\bm{E}_i = \begin{pmatrix}1 & 0 \\ 0 & \sigma_2 \end{pmatrix}  (\bm{Z}_i - \bm{V}_i \bm{\delta}) ,
\end{eqnarray}
and we have $\bm{E}_i| \left( q_i, \bm{\delta}, \nu, \alpha \right) \sim \bm{N}_2( \bm{0}_2, \alpha \bm{\Sigma}/q_i).$
Since the prior $\bm{\Sigma}\sim \bm{W}_2^{-1}(\nu_0, \bm{I}_2)$ implies the priors in (\ref{eq::prior_omega}) and (\ref{eq::prior_sigma2}),
we first draw $ \sigma_2^2|\bm{\Omega} \sim  \{ (1- \rho^2)\chi^2_{\nu_0} \} ^{-1}    $, and then transform the data to get $\bm{E}_i$ using (\ref{eq::trans}). 
The conditional posterior of $\bm{\Sigma}$ is
\begin{eqnarray*}
&&  |  \bm{\Sigma} |^{-(\nu_0 + 3)/2}   \exp\left\{ -\frac{1}{2} \text{tr}( \bm{\Sigma}^{-1} )  \right\}    \prod_{i=1}^N  | \alpha \bm{\Sigma}/ q_i |^{-1/2} \exp\left\{   -\frac{1}{2\alpha} \sum_{i=1}^N q_i \bm{E}_i^\top \bm{\Sigma}^{-1} \bm{E}_i   \right\}    \\
&\propto&  |  \bm{\Sigma}   |^{-(N + \nu_0 + 3)/2}  \exp\left\{    -\frac{1}{2} \text{tr}\left[ \bm{\Sigma}^{-1} (\bm{S}  + \bm{I}_2) \right]    \right\}      \\
&\sim &  \bm{W}_2^{-1}( N+\nu_0, \bm{S} + \bm{I}_2 ) ,
\end{eqnarray*}
where $\bm{S} = \sum_{i=1}^N q_i \bm{E}_i \bm{E}_i^\top/\alpha .$
Therefore, we draw $\bm{\Sigma}  | \left(   \{\bm{E}_i, q_i\},  \bm{\delta} ,  \nu, \alpha \right)  \sim \bm{W}_2^{-1}( N+\nu_0, \bm{S} + \bm{I}_2 )$, and transform $\bm{\Sigma}$ to 
\begin{eqnarray}\label{eq::trans_b}
\sigma_2^2 = \sigma_{22} \text{ and } \bm{\Omega} = 
\begin{pmatrix} 1 & 0 \\ 0& 1/\sigma_2 \end{pmatrix}  
\bm{\Sigma}  
\begin{pmatrix} 1 & 0 \\ 0& 1/\sigma_2 \end{pmatrix} .
\end{eqnarray}

Given $(\{\bm{Z}_i, \bm{V}_i, q_i\}, \bm{\delta}, \bm{\Omega}, \alpha)$, the conditional posterior density of $\nu$ is:
\begin{eqnarray}\label{eq::p_nu}
f(\nu|  \{  \bm{Z}_i,  \bm{V}_i ,q_i \}, \bm{\delta}, \bm{\Omega}, \alpha ) \propto 
\exp\left\{    N\nu \log(\nu/2) /2  - N\log \Gamma(\nu/2)  + (\alpha_0 - 1)\log \nu - \xi \nu    \right\},
\end{eqnarray}
where 
$$\xi = \beta_0 + N\log \alpha/2 + \sum_{i=1}^N q_i/(2\alpha)  - \sum_{i=1}^N \log q_i/2.
$$
Unfortunately, the conditional distribution of $\nu$ is not standard.
Geweke (1992) proposed a rejection sampling method using an exponential distribution as a proposal density.
Albert and Chib (1993) were interested in the posterior probabilities for $\nu$ in a finite set, and they suggested sampling $\nu$ from a discrete distribution. 
In our studies, we treat $\nu$ as a continuous parameter as Geweke (1992). We advocate a more accurate Gamma approximation, and its shape parameter $\alpha^*$ and rate parameter $\beta^*$ are discussed in Appendix C. 
The approximate Gamma distribution is a proposal density for the Metropolized Independence Sampler (Liu 2001) for $\nu$, which is a special case of the Metropolis-Hastings Algorithm.
In our simulation studies and real examples, the Gamma approximation works fairly well with acceptance rates higher than $0.95$.
According to Liu (2001), the efficiency of the Metropolized Independence Sampler depends on how close the proposal density is to the target density. Therefore, the optimal acceptance rates of the Metropolized Independence Sampler are usually higher than the random walk Metropolis algorithm.

Algorithm \ref{alg::post_t} summarizes the posterior step for the selection-t model.
\begin{algorithm}
\caption{Posterior step for the selection-t model}
\label{alg::post_t}
\begin{algorithmic}
\item[P-1] 
Draw $\alpha$ according to (\ref{eq::alpha});

\item[P-2] 
Draw $ \bm{\delta} \sim \bm{N}_{K+L}( \widehat{\bm{\mu}}_{\bm{\delta}},   \widehat{\bm{\Sigma}}_{\bm{\delta}} ) $ 
according to (\ref{eq::delta});

\item[P-3] 
Draw $\sigma_2^2$ from its prior $  \{ (1- \rho^2)\chi^2_{\nu_0} \} ^{-1}  $, make the transformation (\ref{eq::trans}), draw $ \bm{\Sigma} \sim  \bm{W}_2^{-1}( N+\nu_0, \bm{S} + \bm{I}_2 )$, and transform to $\bm{\Omega}$ according to (\ref{eq::trans_b});

\item[P-4] 
Given the old value $\nu$, draw a proposal $\nu'\sim \text{Gamma}(\alpha^*, \beta^*)$, with acceptance probability
\begin{eqnarray*}
\min\left\{    1, \frac{ f(\nu' |\cdot) / \text{dgamma}(\nu', \alpha^*, \beta^*)  }{ f(\nu| \cdot   )/ \text{dgamma}(\nu, \alpha^*, \beta^*)  }     \right\},
\end{eqnarray*}
where $f(\nu|\cdot)$ is the density defined in (\ref{eq::p_nu}), and $\text{dgamma}(\nu, \alpha^*, \beta^*)$ is the Gamma density evaluated at $\nu$.

\end{algorithmic}
\end{algorithm}


\section{The Selection-Robit Model}\label{sec::robit}

In this section, we will discuss the selection-t model with binary outcomes, which can be easily generalized to other types of limited dependent outcomes.
We assume that the regression models for the outcome and the selection mechanism are the same as (\ref{eq::outcome}) and (\ref{eq::selection}), but the observed outcome is $y_i = I(y_i^* > 0)$ if $u_i = 1$, and $y_i = NA$ if $u_i = 0$.
In order to get full identification, we assume that $\varepsilon_i$ and $\eta_i$ follow
\begin{eqnarray*}
\begin{pmatrix}
\varepsilon_i\\
\eta_i
\end{pmatrix}
\sim \bm{t}_2 (
\bm{0}_2, \bm{R},
\nu
),
\end{eqnarray*}
with a Normal mixture representation
\begin{eqnarray*}
\begin{pmatrix}
\varepsilon_i\\
\eta_i
\end{pmatrix}
\sim  \bm{ N}_2(\bm{0}_2, 
\alpha 
\bm{R}/ q_i ), \text{ where } q_i\sim \alpha \chi^2_{\nu}/ \nu, i = 1, \cdots, N;
\text{ and } \bm{R} = \begin{pmatrix} 1 & \rho\\ \rho & 1\end{pmatrix} .
\end{eqnarray*}
Since we can only observe the signs of the latent outcome and selection mechanism, the variances of the error terms are restricted to be one for full identification.

A regression model with binary outcomes can be represented as a latent linear model, with only the signs of the latent outcomes observed (Albert and Chib 1993). 
Different distributions of the error terms correspond to different generalized linear models, including Normal (Probit model), logistic (Logit model) and t (Robit model; Liu 2004).
Therefore, we call the selection model with binary outcomes ``selection-Probit'' model and selection-t model with binary outcomes  ``selection-Robit'' model.

For the selection-Robit model, 
direct sampling the correlation matrix $\bm{R}$ involves non-standard distributions.
Again, we solve this problem by parameter expansion. 
The Inverse-Wishart $\bm{W}_2^{-1}(\nu_0, \bm{I}_2)$ prior for $ \bm{\Sigma} = \text{diag}\{\sigma_1, \sigma_2\} ~ \bm{R} ~ \text{diag}\{\sigma_1, \sigma_2\} $ is equivalent to the priors for $(\bm{R},  \sigma_1^2, \sigma_2^2)$:
\begin{eqnarray}
f(\rho) &\propto& (1 - \rho^2)^{-(\nu_0 - 3)/2},  \label{eq::prior_rho}\\
\text{ and } \sigma_j^2|\rho &\sim& \{ (1 - \rho^2) \chi^2_{\nu_0}  \}^{-1} (j=1, 2). \label{eq::prior_sigmas}
\end{eqnarray}
Details of the derivations of (\ref{eq::prior_rho}) of (\ref{eq::prior_sigmas}) are in Appendix B. If we choose $\nu_0 = 3$, the prior distribution for $\rho$ is Uniform$(-1, 1)$. Therefore, the prior $\bm{W}_2^{-1}(3, \bm{I}_2)$ for $\bm{\Sigma}$ is a marginally uniform prior (Barnard, McCulloch and Meng 2000).

For selection-Robit model, the imputation of $\bm{Z}_i$ changes slightly.
The conditional means and variances of $u_i^*$ and $y_i^*$ have the following forms:
\begin{eqnarray*}
&\tilde{\mu}_{u|y} = \bm{w}_i^\top \bm{\gamma}  +  \rho    (y_i^* - \bm{x}_i^\top \bm{\beta}),
&\tilde{\sigma}^2_{u|y} = \alpha(1 - \rho^2)/q_i,\\
&\tilde{\mu}_{y|u} = \bm{x}_i^\top \bm{\beta} + \rho (u_i^* - \bm{w}_i^\top \bm{\gamma}  ),
&\tilde{\sigma}^2_{y|u} = \alpha(1 - \rho^2)/q_i.
\end{eqnarray*}

If $u_i = 1$, we draw the truncated bivariate Normal distribution using the Gibbs sampler:
\begin{eqnarray}
u_i^*| \left(   y_i^*, q_i, y_i, u_i,\bm{\delta}, \rho, \nu, \alpha \right)  &\sim& TN( \tilde{\mu}_{u|y}, \tilde{\sigma}^2_{u|y}; 0, \infty ) , \label{eq::u|y1}\\
y_i^*| \left(    u_i^*, q_i, y_i, u_i,\bm{\delta}, \rho, \nu, \alpha  \right)  &\sim& y_i TN( \tilde{\mu}_{y|u},   \tilde{\sigma}^2_{y|u}; 0, \infty) + (1 - y_i) TN( \tilde{\mu}_{y|u},   \tilde{\sigma}^2_{y|u}; -\infty, 0).\nonumber \\
&&\label{eq::y|u1}
\end{eqnarray}
If $u_i = 0$ and $y_i$ is missing, we impute the missing data by two steps:
\begin{eqnarray} 
u_i^*| \left(  q_i, y_i, u_i,\bm{\delta}, \rho, \nu, \alpha \right)  &\sim& TN( \bm{w}_i^\top \bm{\gamma}, \alpha/q_i; -\infty, 0 ),
\label{eq::u0}\\
 y_i^*|  \left( u_i^*, q_i, y_i, u_i,\bm{\delta}, \rho, \nu, \alpha \right)  &\sim& N(   \tilde{\mu}_{y|u},   \tilde{\sigma}^2_{y|u}  ).\label{eq::y|u0}
\end{eqnarray}

The full conditional distributions of $\bm{\delta}$ and $\nu$ have the same forms as in the previous section, except that $\bm{\Omega}$ is replaced by $\bm{R}$.
The full conditional distribution of $\bm{R}$ does not have conjugate form,
which can be circumvented by parameter expansion.
Define
\begin{eqnarray}\label{eq::trans12}
\tilde{ \bm{E} } _i = \begin{pmatrix}\sigma_1 & 0 \\ 0 & \sigma_2 \end{pmatrix}  (\bm{Z}_i - \bm{V}_i \bm{\delta}),
\end{eqnarray}
and we have $\tilde{ \bm{E} } _i |\left(  q_i, \bm{\delta}, \nu, \alpha \right)  \sim \bm{N}_2( \bm{0}_2, \alpha \bm{\Sigma}/q_i).$
The prior $\bm{\Sigma}\sim \bm{W}_2^{-1}(\nu_0, \bm{I}_2)$ implies prior for $(\bm{R}, \sigma_1^2, \sigma_2^2)$ 
as shown in (\ref{eq::prior_rho}) and (\ref{eq::prior_sigmas}).
We first independently draw $ \sigma_j^2|\bm{\Omega} \sim  \{ (1- \rho^2)\chi^2_{\nu_0} \} ^{-1}   (j=1,2) $, and then transform the data to obtain $ \tilde{ \bm{E} } _i   $ using (\ref{eq::trans12}). 
The conditional posterior of $\bm{\Sigma}$ is
$ \bm{W}_2^{-1}( N+\nu_0, \tilde{ \bm{S} } + \bm{I}_2 ),$ where $\tilde{ \bm{S} }  = \sum_{i=1}^N q_i \tilde{ \bm{E}}_i \tilde{ \bm{E}} _i^\top/\alpha .$
Therefore, we draw $\bm{ \Sigma}  |  (  \{ \tilde{ \bm{E} } _i, q_i\},  \bm{\delta} ,  \nu, \alpha ) \sim \bm{W}_2^{-1}( N+\nu_0, \tilde{ \bm{S} }  + \bm{I}_2 )$, and transform $\bm{\Sigma}$ to 
\begin{eqnarray}\label{eq::trans_b12}
\sigma_1^2 = \sigma_{11},
\sigma_2^2 = \sigma_{22} \text{ and } \bm{R} = 
\begin{pmatrix} 1/\sigma_1 & 0 \\ 0& 1/\sigma_2 \end{pmatrix}  
\Sigma  
\begin{pmatrix} 1/\sigma_1 & 0 \\ 0& 1/\sigma_2 \end{pmatrix} .
\end{eqnarray}

Algorithm \ref{alg::impute_tobit} and  \ref{alg::post_tobit} summarize the imputation and posterior steps for the selection-Robit model, respectively.
\begin{algorithm}
\caption{Imputation step for the selection-Robit model}
\label{alg::impute_tobit}
\begin{algorithmic}
\item[I$'$-1] Draw $\alpha\sim b/\chi^2_c$;

\item[I$'$-2] Impute $(y_i^*, u_i^*)$ according to (\ref{eq::u|y1}) to (\ref{eq::y|u1});

\item[I$'$-3] Draw $q_i$ the same as I-3 except that $\bm{\Omega}$ is replaced by $\bm{R}$.
\end{algorithmic}
\end{algorithm}

\begin{algorithm}
\caption{Imputation step for the selection-Robit model}
\label{alg::post_tobit}
\begin{algorithmic}
\item[P$'$-1] 
Draw $\alpha$ the same as P-1 except that $\bm{\Omega}$ is replaced by $\bm{R}$;

\item[P$'$-2] 
Draw $\bm{\delta}$ the same as P-2 except that $\bm{\Omega}$ is replaced by $\bm{R}$; 

\item[P$'$-3] 
Draw $\sigma_1^2$ and $\sigma_2^2$ independently from their priors $\{(1 - \rho^2)\chi^2_{\nu_0}\}^{-1}$, make the transformation (\ref{eq::trans12}), draw $\bm{\Sigma}\sim \bm{W}_2^{-1}( N+\nu_0, \tilde{ \bm{S} }  + \bm{I}_2 )$, and transform $\bm{\Sigma}$ back to $\bm{R}$ according to (\ref{eq::trans_b12});

\item[P$'$-4]
Draw $\nu$ the same as P-4.
 
\end{algorithmic}
\end{algorithm}


\section{Simulation Studies} \label{sec::simulation}
In order to evaluate the finite sample properties of the new Bayesian procedures,
several simulation studies are presented in this section.
The MCMC algorithms 1 to 4 for the selection-t model and selection-Robit model can be easily refined to be the algorithms for the selection model and selection-Probit model, by restricting $\alpha =1$, $\nu = +\infty$ and $q_i = 1(i=1,\cdots,N)$.
Throughout our simulation studies and our empirical studies, the parameters for the prior distributions are chosen as follows: $\bm{\mu}_0 = \bm{0}_{K+L}, \bm{\Sigma}_0 = \text{diag}\{1, \cdots,  1\}/100, \nu_0=3, \alpha_0 = 1, \beta_0=0.1, b = 0.1$, and $c = 0.1$. The results are not sensitive to other choices of hyperparameters.
We run the MCMC algorithms for $5\times 10^4$ iterations, with the first $10^4$ draws discarded as a burn in period.
The results from multiple chains differ very slightly and all of them converge with Gelman-Rubin diagnostic statistics close to $1$, and we only present the result from a single chain in each of our simulation study and real application.

\subsection{Data Generated from Models with Normal Errors}\label{subsec::normal}
\begin{figure}[ht]
\begin{tabular}{p{\columnwidth}}
\includegraphics[width = \textwidth]{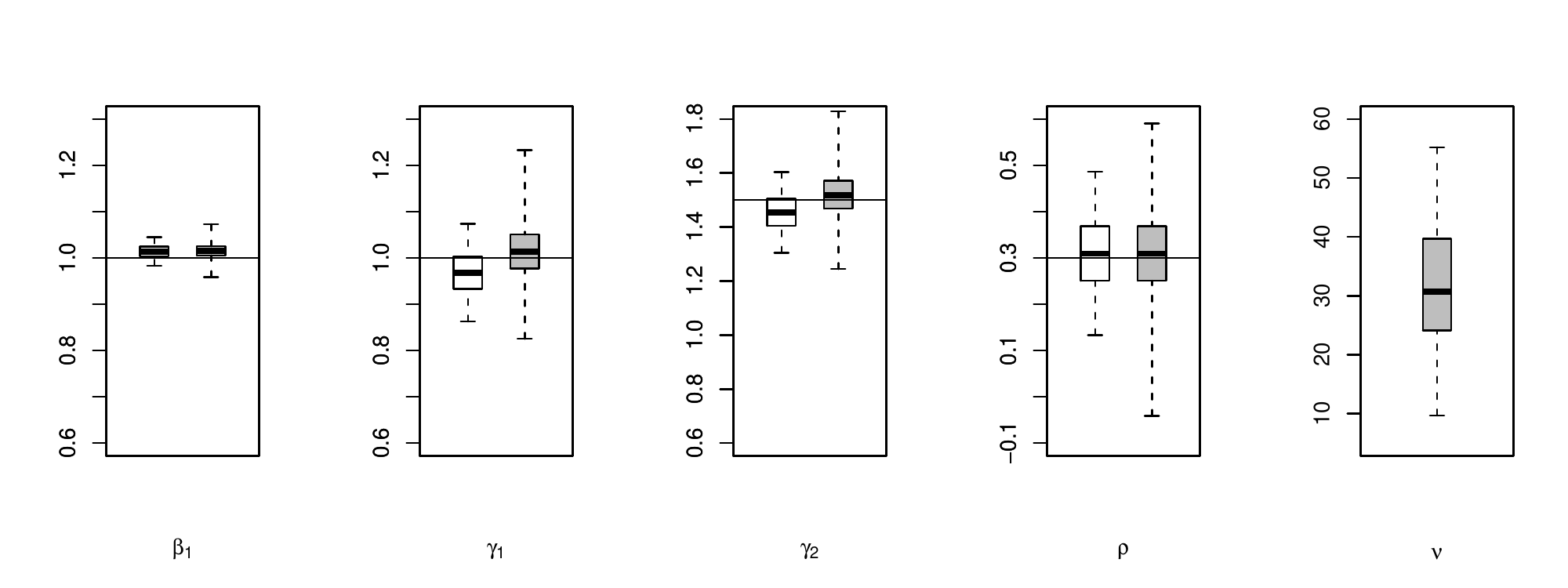}
\\
(a) Data generated from the selection model with Normal errors and analyzed by the selection model (white) and the Selection-t model (grey).\\
\includegraphics[width = \textwidth]{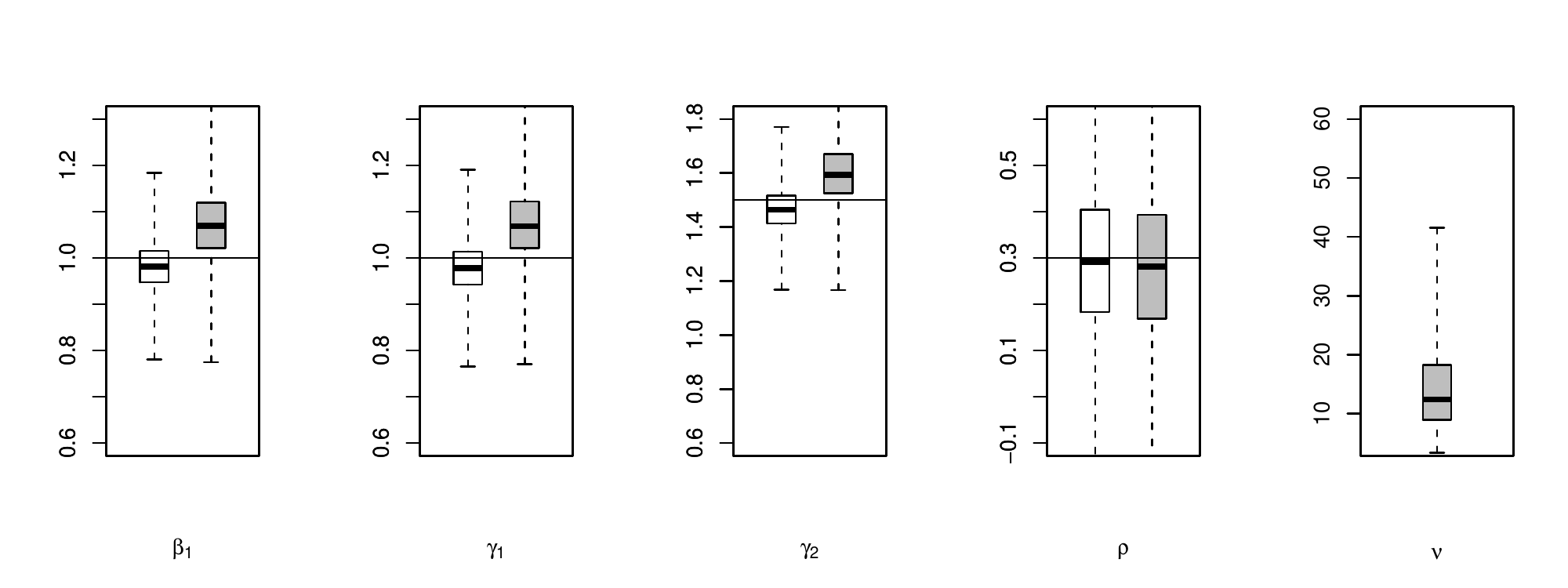}
\\
(b) Data generated from the Selection-Probit model and analyzed by Selection-Probit model (white) and Selection-Robit model (grey).
\end{tabular}
\caption{Selection Models with Normal Errors}\label{fg::normal}
\end{figure}

Selection-t models are more general than the selection models, and they converge to the selection models when $\nu\rightarrow +\infty$. 
We first generate the observed data from a selection model with bivariate Normal errors.
We generate the covariates from $x_i \sim N(0, 2^2), w_i\sim N(0, 2^2)$, and $x_i$ is independent of $w_i$; and generate the latent outcome and selection mechanism from $y_i^* = 0.5 + x_i + \varepsilon_i, u_i^* = 2 + x_i + 1.5 w_i + \eta_i$, and 
$$
\begin{pmatrix}  \varepsilon_i \\ \eta_i \end{pmatrix}
\sim
\bm{N}_2 \left[   
\begin{pmatrix}  0\\ 0 \end{pmatrix},
\bm{\Omega}_0 = 
\begin{pmatrix} 1& 0.3\\0.3& 1 \end{pmatrix}
\right].
$$
The selection indicator is $u_i = I(u_i^*>0)$, and the outcome is $y_i = y_i^*$ if $u_i=1$ and $y_i = \text{NA}$ if $u_i=0.$
In our generated data set, about $30\%$ outcomes are missing.
We apply the Bayesian procedures for both the selection model and selection-t model, and the posterior distributions of the parameters are summarized in the boxplots in Figure \ref{fg::normal}(a). The boxplots in white and grey are obtained under the selection model and selection-t model, respectively. The posterior distributions of the parameters $(\beta_1, \gamma_1, \gamma_2, \rho)$ concentrate near their true values under both models. And the posterior draws of $\nu$ under the selection-t model take very large values.
It is known that t distributions with large degrees of freedom approximate the Normal distribution. Therefore, large values of the posterior draws of $\nu$ are evidence of Normality.

Replacing $y_i$ with $I(y_i > 0 )$ in the same data set, we
implement the Bayesian procedures for both the selection-Probit and the selection-Robit model.
The posterior distributions of the parameters are summarized in the boxplots in Figure \ref{fg::normal}(b), with the boxplots in white for the selection-Probit model and the boxplots in grey for the selection-Robit model.
Although more diffused than the distributions in Figure \ref{fg::normal}(a) due to loss of information after dichotomizing the continuous outcomes, the posterior distributions of the parameters in Figure \ref{fg::normal}(b) concentrate near the true parameters. And the large values of posterior draws of $\nu$ under the selection-Robit model show evidence of Normality.

\subsection{Data Generated from Models with t Errors}\label{subsec::t3}

\begin{figure}[ht]
\begin{tabular}{p{\columnwidth}}
\includegraphics[width = \textwidth]{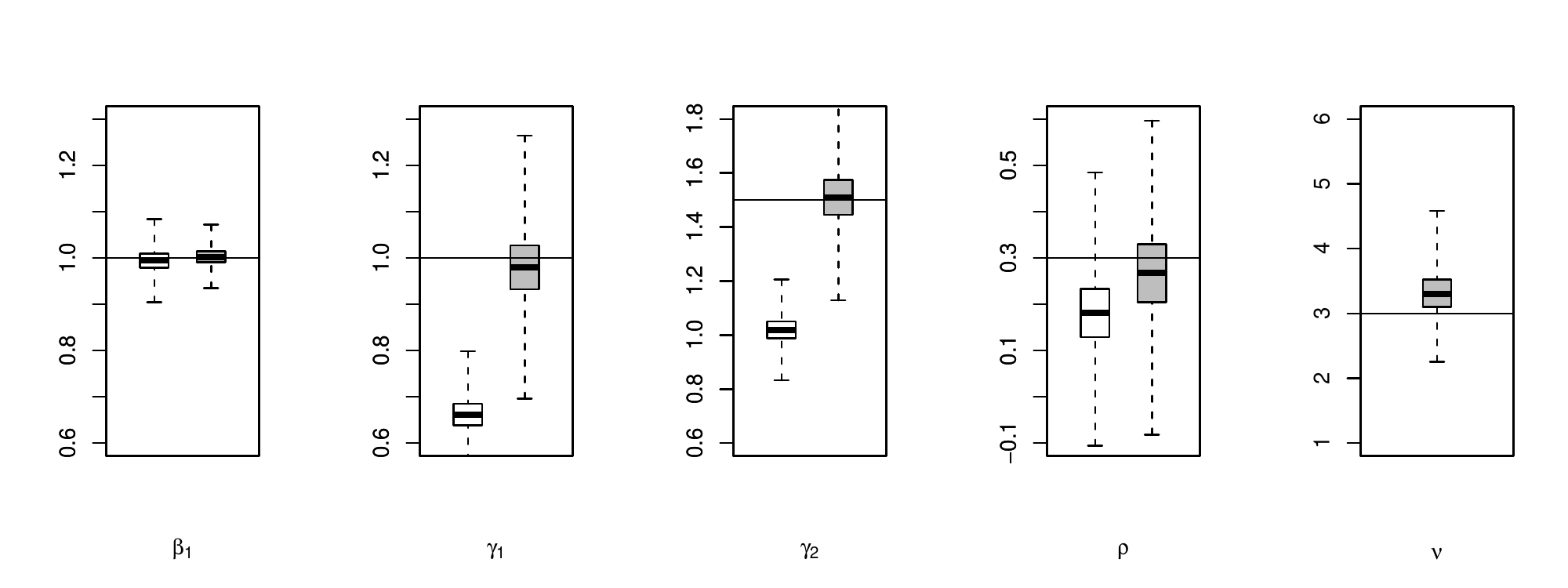}
\\
(a) Data generated from the Selection-t model with Normal errors and analyzed by the selection model (white) and the Selection-t model (grey).\\
\includegraphics[width = \textwidth]{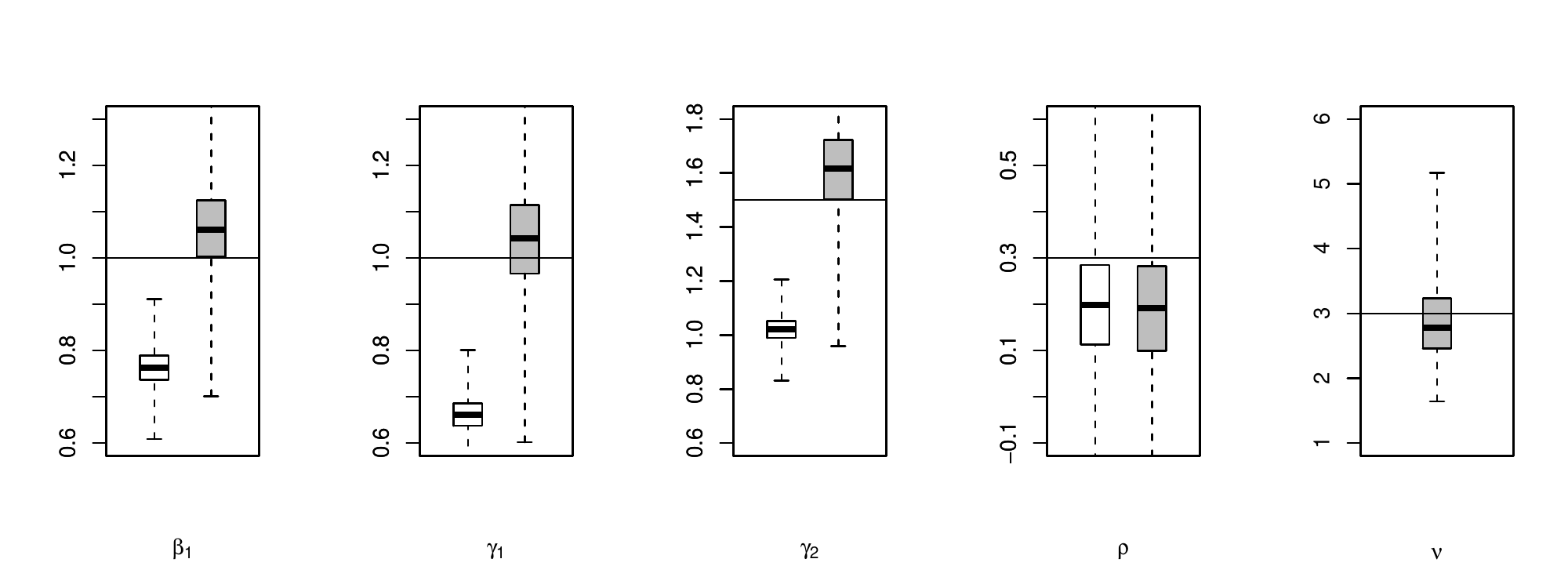}
\\
(b) Data generated from the Selection-Robit model and analyzed by Selection-Probit model (white) and Selection-Robit model (grey).
\end{tabular}
\caption{Selection Models with t Errors}\label{fg::t3}
\end{figure}

We generate data from a selection-t model. The data generating process is the same as Section \ref{subsec::normal} except that the bivariate Normal distribution is replaced by a bivariate t distribution with degrees of freedom $\nu=3$. 
In our generated data set, about $30\%$ outcomes are missing.
The posterior distributions of the parameters are summarized in the boxplots in Figure \ref{fg::t3}(a) under both the selection model (in white) and the selection-t model (in grey). The posterior distributions under the selection-t model are close to the true parameters, but those under the selection model are far from the true parameters (e.g., $\gamma_1$ and $\gamma_2$ in Figure \ref{fg::t3}(a)).
Again, replacing $y_i$ with $I(y_i>0)$, we apply both the Bayesian procedures for the selection-Probit model and the selection-Robit model, and the boxplots of the posterior distributions are shown in Figure \ref{fg::t3}(b). We can see that the posterior distributions of $(\beta_1, \gamma_1, \gamma_2)$ under the selection-Probit model are farther from the true parameters than those under the selection-Robit model.

\subsection{Data Generated from Models with Gaussian Mixture Errors}

\begin{figure}[ht]
\begin{tabular}{p{\columnwidth}}
\includegraphics[width = \textwidth]{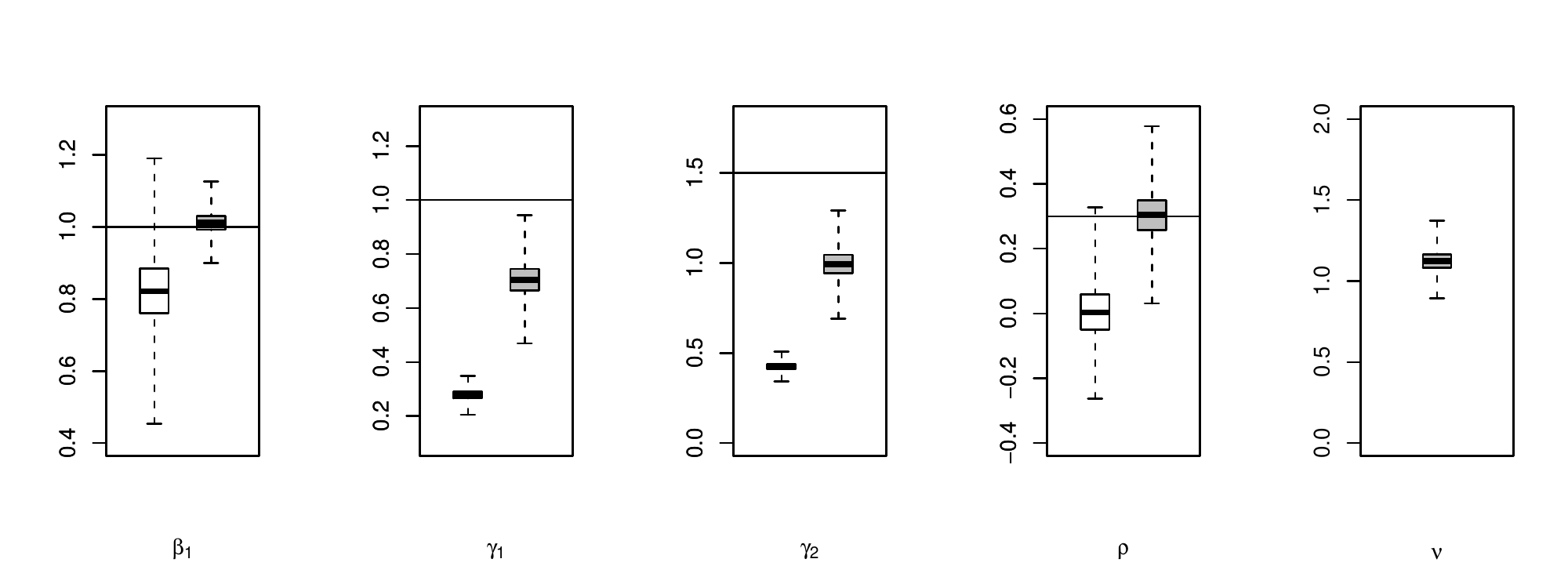}
\\
(a) Data generated from the Selection model with Gaussian Mixture errors and analyzed by the selection model (white) and the Selection-t model (grey).\\
\includegraphics[width = \textwidth]{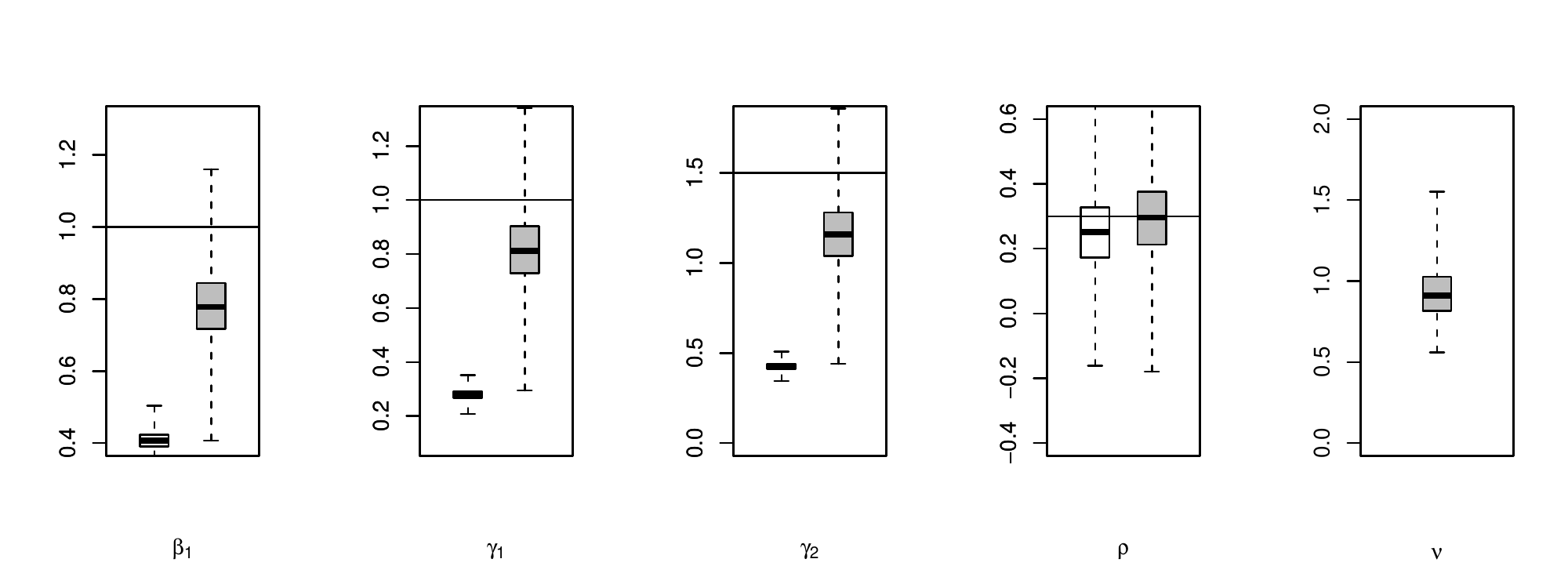}
\\
(b) Data generated from the Selection model with Gaussian Mixture errors and analyzed by Selection-Probit model (white) and Selection-Robit model (grey).
\end{tabular}
\caption{Selection Models with Gaussian Mixture Errors}\label{fg::mix}
\end{figure}

We generate data from a selection model with Gaussian Mixture errors: 
$$
0.4
\bm{N}_2(\bm{0}_2, \bm{\Omega}_0)+
0.3\bm{N}_2(\bm{0}_2, 2 \bm{\Omega}_0)+
0.2 \bm{N}_2(\bm{0}_2, 4\bm{\Omega}_0)+
0.1 \bm{N}_2(\bm{0}_2,8 \bm{\Omega}_0)+
0.1\bm{N}_2(\bm{0}_2, 16\bm{\Omega}_0),
$$
where $ \bm{\Omega}_0$ is defined in Section \ref{subsec::normal}.
In our generated data set, about $33\%$ of the outcomes are missing.
Figure \ref{fg::mix}(a) are the boxplots of the posterior distributions of the parameters under the selection model (in white) and selection-t model (in grey),
and Figure \ref{fg::mix}(b) are the boxplots of the posterior distributions of the parameters under the selection-Probit (in white) and selection-Robit model (in grey).
%
%
Since both the selection models based on Normal and t distribution are misspecified, most posterior distributions do not concentrate very near the true parameters. The behaviors of the selection and selection-Probit models are very wild (e.g. $\gamma_1$ and $\gamma_2$) due to the heavy-tails generated by the Gaussian Mixture errors, while the behaviors of the selection-t and selection-Robit model are much more robust to the model misspecifications.

\section{Empirical Studies}\label{sec::applications}

\subsection{Ambulatory Expenditures}
We apply our Bayesian procedures to the data about ambulatory expenditures. The data are taken from Cameron and Trivedi (2010), which were re-analyzed by Marchenko and Genton (2012) using frequentists' procedure for the selection-t model.
The data can be downloaded from \url{http://cameron.econ.ucdavis.edu/musbook/mus.html}.
In our analysis, we choose log expenditures (\texttt{lambexp}) as the outcome variable $y$. The covariates in the outcome equation are $\bm{x} = (1, \texttt{age}, \texttt{female}, \texttt{educ}, \texttt{blhisp}, \texttt{totchr}, \texttt{ins})$, including age, gender, education status, ethnicity, number of chronic diseases and insurance status.
The exclusion restriction assumption holds by including the income variable into the selection equation, i.e.,
$\bm{w} = (\bm{x}, \texttt{income})$.
In order to compare with the frequentists' approach, we choose the same set of covariates as Marchenko and Genton (2012).
We would expect similar results from both frequentists' and Bayesian procedures, because the prior information will be overwhelmed by the data with relatively large sample size ($n=3328$ with $526$ missing outcomes).
Table \ref{tb::mus16} shows that the results of frequentists' and Bayesian methods under the same model are very close to each other, while the selection model and the selection-t model give us different results.
The consistent results from both the frequentists' procedure and the Bayesian procedure verify our MCMC code and convergence of the Markov Chains indirectly.

\newsavebox{\tablebox}
\begin{lrbox}{\tablebox}
\begin{tabular}{|  rrrrr rrrr   |}
\hline
 &  \multicolumn{2}{c}{ Selection} &  \multicolumn{2}{c}{ Selection-t } & \multicolumn{2}{c}{ Selection (Bayes)} &  \multicolumn{2}{c|}{Selection-t (Bayes)} \\
\hline
\multicolumn{9}{|l|}{Outcome Model} \\
age            & $ 0.212 $&${(0.167, 0.257)} $ & $ 0.207 $&${(0.163, 0.251)}$ & $ 0.211 $&${(0.165, 0.256) }$& $0.207 $&${(0.162, 0.251) }$\\
female       & $0.348 $&${(0.230, 0.466)  }$&  $ 0.307 $&${(0.196, 0.417)}$ & $0.339 $&${(0.217, 0.456) }$&  $0.306 $&${(0.194, 0.417) }$\\
educ          & $0.019 $&${(-0.002, 0.039) }$ & $0.017 $&${(-0.003, 0.037)}$ & $0.018 $&${(-0.003, 0.037) }$&  $0.017$&${ (-0.003, 0.038)}$\\
blhisp        & $-0.219 $&${(-0.336, -0.102)}$ & $-0.193 $&${(-0.306, -0.080)}$ &$ -0.213 $&${(-0.329, -0.094)}$ & $-0.193 $&${(-0.306, -0.078)}$\\
totchr         & $ 0.540 $&${(0.463, 0.617) }$ &  $0.513 $&${(0.443, 0.583) }$&  $0.534 $&${(0.453, 0.611) }$&  $0.512$&${ (0.441, 0.583)}$\\
ins             &  $-0.030 $&${(-0.130, 0.070 )}$ & $-0.053 $&${(-0.151, 0.046) }$&$ -0.033$&${ (-0.133, 0.068)}$ &$ -0.054 $&${(-0.153, 0.046)}$ \\
\hline
\multicolumn{9}{|l|}{Selection Model} \\
age           &  $0.088 $&${(0.034, 0.142) }$  &   $0.099 $&${(0.040, 0.157)}$ &   $0.088 $&${(0.033, 0.142) }$& $ 0.099 $&${(0.040, 0.157) }$\\
female       &  $0.663 $&${(0.543, 0.782) }$ &  $0.725 $&${(0.591, 0.859) }$&   $0.664 $&${(0.544, 0.784) }$& $0.729 $&${(0.597, 0.867)}$\\
educ          &  $0.062 $&${(0.038, 0.086)}$ &   $0.065 $&${(0.040, 0.090) }$&   $0.062 $&${(0.038, 0.085) }$& $0.065 $&${(0.040, 0.090)}$\\
blhisp         &  $-0.364$&${ (-0.485, -0.243)}$ & $-0.394 $&${(-0.524, -0.263)}$ &$ -0.364 $&${(-0.485, -0.244)}$& $-0.394$&${ (-0.525, -0.265)}$\\
totchr         & $0.797 $&${(0.658, 0.936) }$ &  $0.890 $&${(0.719, 1.061)}$ &   $0.795 $&${(0.660, 0.936) }$& $0.893 $&${(0.733, 1.075)}$\\
ins             & $ 0.170 $&${(0.047, 0.293) }$ & $0.180 $&${(0.048, 0.313)}$ &   $0.169 $&${(0.045, 0.291) }$& $0.180 $&${(0.047, 0.314)}$\\
income      & $ 0.003 $&${(0.000, 0.005) }$ & $ 0.003 $&${(0.000, 0.006)}$ &   $0.003 $&${(0.000, 0.005)}$ & $0.003 $&${(0.000, 0.006)}$\\
\hline
$\sigma$   & $1.271 $&${(1.236, 1.308)} $&  $1.195 $&${(1.146, 1.246)}$ & $1.277 $&${(1.241, 1.324) }$& $1.195 $&${(1.148, 1.249)}$\\
$\rho$       &  $-0.131$&${ (-0.401, 0.161)}$ &  $-0.322 $&${(-0.526, -0.083)}$ &  $ -0.159 $&${(-0.462, 0.108) }$& $-0.327 $&${(-0.536, -0.085)}$\\
$\nu$        & $+\infty$                     &    &  $12.938 $&${(8.391, 19.917)}$ &   $+\infty$& & $12.913$&${ (8.841, 22.447)}$\\
\hline
\end{tabular}
\end{lrbox}

\begin{table}[ht]
\caption{The Ambulatory Expenditures Example}\label{tb::mus16}
\begin{center}
\resizebox{\textwidth}{!}{\usebox{\tablebox}}
\end{center}
{\small Note: MLEs and $95\%$ confidence intervals of the selection model (column 2 and 3), the selection-t model (column 4 and 5);
Bayesian posterior means and $95\%$ credible intervals of the selection model (column 6 and 7), the selection-t model (column 8 and 9). 
}
\end{table}

The value of $\rho$ measures the sample selection effect, with $\rho=0$ indicating the absence of the sample selection bias. Under the selection model, both the frequentists' $95\%$ confidence interval and the Bayesian $95\%$ posterior credible interval of $\rho$ contain zero, which indicates weak evidence of the sample selection bias.
However, under the selection-t model, neither the frequentists' $95\%$ confidence interval nor the Bayesian $95\%$ posterior credible interval of $\rho$ contains zero, which suggests the existence of sample selection effect.
Different inferences on the existence of selection effect give us different statistical and economic interpretations. Absence of selection effect ($\rho=0$) implies that the outcomes are missing at random (Little and Rubin, 2002), and the observed outcomes are representative for inference of the ambulatory expenditures given the observed covariates. 
The results of the selection model and the selection-t model are different, because of the heavy-tailedness in the data.
Figure \ref{fg::nu}(a) depicts the posterior distribution of $\nu$ with 95\% posterior credible interval $(8.841, 22.447)$, which is evidence of heavy-tailedness.

%
%

\subsection{Wage Offer Function for Married Women}
We re-analyze the data from Mroz (1987) and Wooldridge (2002) to estimate the wage offer function for married women. The data set can be found in the \texttt{R} package \texttt{sampleSelection}.
The outcome of interest is the log of \texttt{wage}, which are missing for $325$ individuals and observed for $428$ individuals.
The covariates in the outcome equation are $\bm{x} = (1, \texttt{educ}, \texttt{exper}, \texttt{exper}^2)$, including education status, experience and its squared term.
The covariates in the selection equation includes other income, age, number of young children and number of older children as additional variables, i.e., $\bm{w} = (\bm{x},  \texttt{nwifeinc}, \texttt{age}, \texttt{kids5}, \texttt{kids618})$.
Wooldridge (2002) used Heckman's two-step procedure, because the MLE in this particular data is numerically unstable.
The \texttt{heckman} procedure in old versions of \texttt{Stata} (e.g. \texttt{Stata} 10) converges to a local maximum.
The results in Table \ref{tb::wage} are from the latest version of \texttt{Stata} (\texttt{Stata} 13), which are the same as the results from \texttt{R} function \texttt{selection()} in the package \texttt{sampleSelection} (\texttt{R} version 2.10.0 with \texttt{sampleSelection}  package version 0.7.2).
Bayesian procedures are more preferable in this case, because they are less sensitive to the initial values.
Table \ref{tb::wage} shows that the Bayesian and frequentists' methods for the selection model give us similar results. 
Most regression coefficients under the selection model and the selection-t model do not differ dramatically. However, there does exist some different economic interpretations under different models.
For instance, according to the results from Bayesian inference, the \texttt{exper} variable enter the outcome equation as $-0.001\text{\texttt{exper}}^2 + 0.043\text{\texttt{exper}}$ and $0.026\text{\texttt{exper}}$ under the selection model and selection-t model, respectively.
The former implies that $\log(\text{\texttt{wage}})$ is a quadratic function of \texttt{exper} with a maximum point at $21.5$, while the latter implies
that it is a linear thus increasing function of \texttt{exper}.

\newsavebox{\tableboxbox}
\begin{lrbox}{\tableboxbox}
\begin{tabular}{|  rrrrrrrrr |}
\hline
        & \multicolumn{2}{c}{ Selection } & \multicolumn{2}{c}{ Selection-t} & \multicolumn{2}{c}{ Selection (Bayes) } & \multicolumn{2}{c|}{ Selection-t (Bayes) } \\
        \hline
        \multicolumn{9}{|l|}{Outcome Model}\\
educ&  $0.108$ &  ${ (0.079, 0.137)}$ &          $0.108$   & ${(0.085, 0.132)}$         &   $0.108 $&${(0.078, 0.137)}$ &  $0.109 $&${(0.085, 0.132)}$\\
exper&  $0.043$ & ${(0.014, 0.072) }$&         $0.025$    &  ${(0.001, 0.050)}$         & $0.043 $&${(0.012, 0.072) }$  & $0.026 $&${(0.001, 0.051)}$\\
exper$^2$& $-0.001$  & ${ (-0.002, 0.000)}$& $ -0.000$ &${(-0.001, 0.002)}$         & $-0.001$&${ (-0.002, 0.000)}$& $-0.000$&${ (-0.001, 0.002)}$\\
\hline
        \multicolumn{9}{|l|}{Selection Model}\\
educ & $0.131$& ${ (0.082, 0.181)}$&             $0.140$&${(0.079, 0.201)}$        & $0.132 $&${(0.083, 0.182)} $& $0.143 $&${(0.083, 0.208)}$\\
exper &$ 0.123$ & ${ (0.087, 0.160)}$ &           $0.150$&${(0.104, 0.195)}$        & $0.124$&${ (0.087, 0.161) }$&  $0.151$&${ (0.105, 0.199)}$\\
exper$^2$ & $-0.002$&  ${ (-0.003, -0.001)}$& $-0.002$&${(-0.004, -0.001)}$       &$-0.002$&${ (-0.003, -0.001)}$ & $-0.002$&${ (-0.004, -0.001)}$\\
nwifeinc & $-0.012$ & ${(-0.022, -0.003)}$ &    $-0.012$&${(-0.024, 0.000)}$       &$-0.012 $&${(-0.022, -0.003)}$ &  $-0.012 $&${(-0.024, -0.001)}$\\
age & $-0.053$ &${ (-0.069, -0.036)}$ &           $-0.064$&${(-0.085, -0.043)}$       &$-0.053$&${ (-0.069, -0.036)}$ & $-0.065 $&${(-0.087, -0.045)}$ \\
kids5 & $-0.867$ &${(-1.100, -0.635)}$ &         $-1.043$&${(-1.339, -0.749)}$       &$-0.870$&${ (-1.101, -0.640)}$ & $-1.060$&${ (-1.374, -0.771)}$\\
kids618 & $0.036$&${ (-0.049, 0.121) }$&       $0.034$&${(-0.070, 0.139)}$      &$0.037$&${ (-0.048, 0.123)}$ & $0.034$&${ (-0.071, 0.142)}$\\
\hline
$\sigma$ & $0.663$&${ (0.619, 0.708)}$ &       $0.444$&${(0.394, 0.500)}$      & $0.670$&${ (0.627, 0.720) }$& $0.451$&${ (0.399, 0.508)}$\\
$\rho$  & $0.027$&$ { (-0.262, 0.315)} $&          $-0.383$&${(-0.632, -0.061)}$     &$0.019$&${ (-0.308, 0.281)}$ & $-0.362$&${ (-0.605, -0.053)}$\\
$\nu$ &$+\infty$ &          &               $3.061$&${(2.340, 4.183)}$      & $+\infty$        &             & $3.094$&${ (2.296, 4.318)}$\\
 \hline
\end{tabular}
\end{lrbox}

\begin{table}[ht]
\caption{The Wage Offer Example}
\label{tb::wage}
\begin{center}
\resizebox{\textwidth}{!}{\usebox{\tableboxbox}}
\end{center}
{\small Note: MLEs and $95\%$ confidence intervals of the selection model (column 2 and 3), the selection-t model (column 4 and 5);
Bayesian posterior means and $95\%$ credible intervals of the selection model (column 6 and 7), the selection-t model (column 8 and 9). 
}
\end{table}

Under the selection model,
the $95\%$ confidence interval and posterior credible interval of $\rho$ cover zero, and the evidence for the sample selection effect is weak. However, the result using the selection-t model differs a lot, because the posterior credible interval of $\rho$ does not cover zero, and it indicates the existence of the sample selection effect. The conclusion changes, because $\nu$ is very small with posterior credible interval $(2.296, 4.318)$ and histogram in Figure \ref{fg::nu}(b). 
The tail behavior of t distribution with small degrees of freedom
is very different from the tail behavior of the normal distribution. 
From the summary statistics of the observed values of the outcome, we find that the kurtosis of the observed outcome is greater than five, which is much greater than three, the kurtosis of normal distributions. Heavy-tailedness problem in this data is very severe, and the selection-t model is more preferable than the selection model.

\subsection{HIV Survey with Nonparticipation}

We apply our Bayesian procedure for selection-Robit model to the data
from the 2007 Zambia Demographic and Health Survey (B\"arnighausen et al., 2011).
We are interested in estimating the HIV prevalence rate among men. However,
in the survey, $6416$ male participants were tested for HIV status, but 
$1318$ males rejected to take the HIV test. For those nonparticipants of the HIV test, their binary outcomes $y_i$ (HIV status)
are missing with $u_i = 0$.
The covariates in the outcome equation are $\bm{x}  = (1,  \texttt{age} , \texttt{income}, \texttt{marital}, \texttt{condom}, \texttt{evertestHIV}, \texttt{smoking} , \texttt{location})$, including age, income, marital status, condom use, ever tested HIV, smoking status, and the location of the individuals. And the covariates in the selection equation are $\bm{w} $ including both $\bm{x} $ and the identity of the interviewer, with the latter included for the exclusion restriction assumption.
In Table \ref{tb::HIV}, we show the MLEs and $95\%$ confidence intervals from the \texttt{heckprob} procedure in \texttt{Stata}, and the Bayesian posterior medians and $95\%$ credible intervals of the selection-Robit model. The Bayesian posterior distributions of the selection-Probit is omitted here, since the extreme draws of $\rho$ from the Gibbs sampler make the algorithm numerically unstable. 
Therefore, we find the numerical stability in presence of heavy-tailedness in the outcome as another advantage of the selection-Robit model.
The frequentists' procedure for the selection-Robit is also omitted, since it is not very straightforward to modify Marchenko and Genton (2012)'s procedure to deal with binary outcomes.

\begin{figure}[b]
\begin{tabular}{ccc}
\includegraphics[width = .33\textwidth]{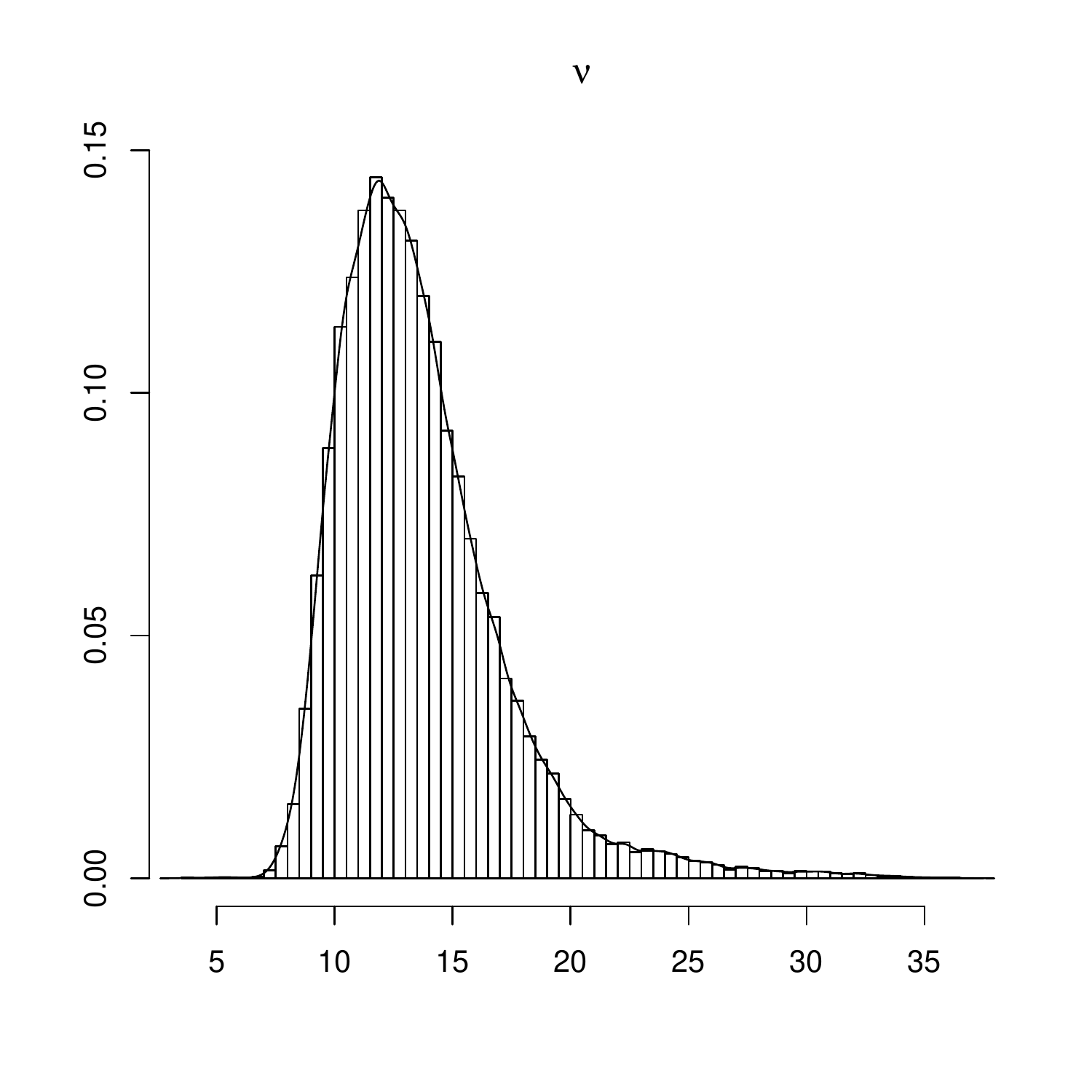}
&
\includegraphics[width = .33\textwidth]{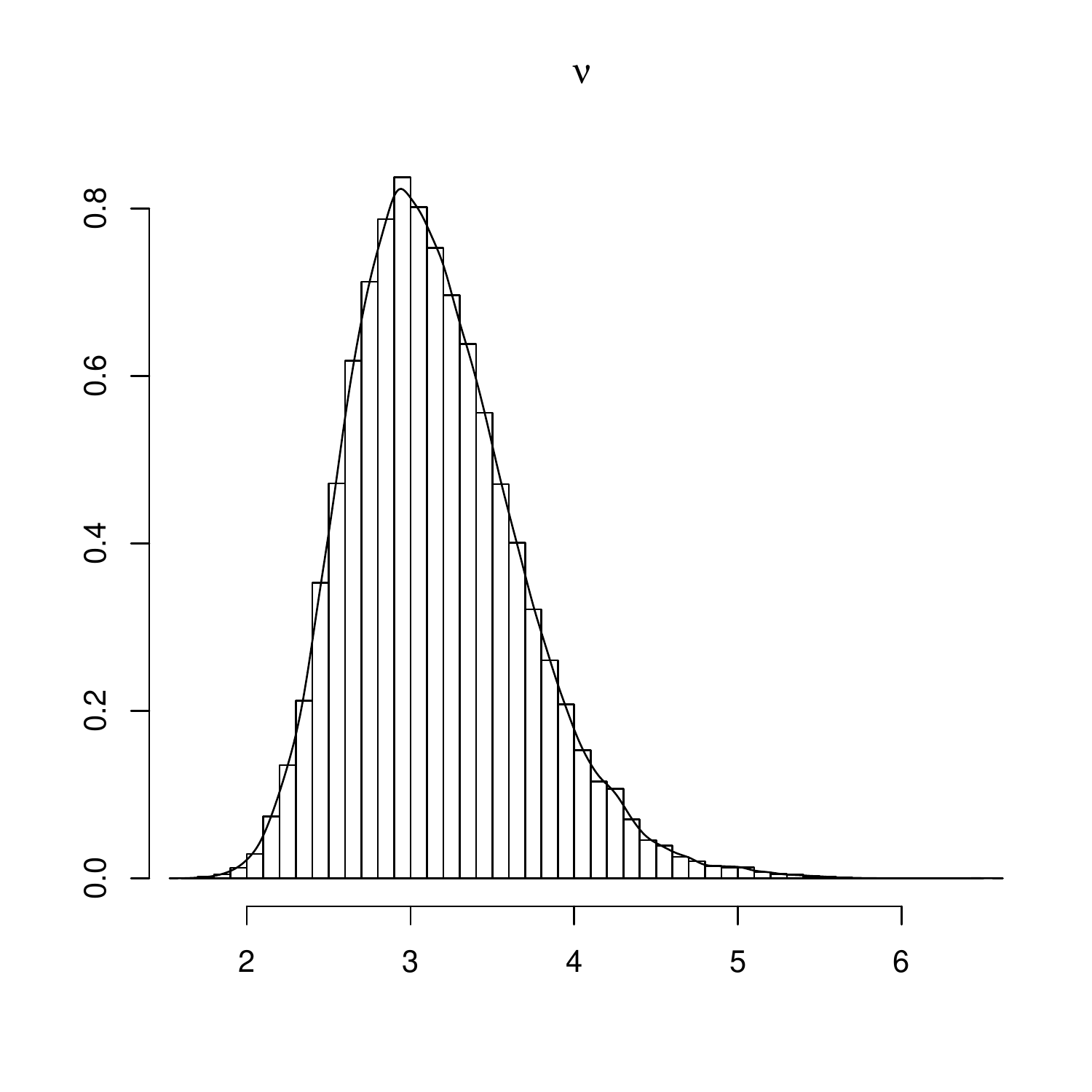} 
&
\includegraphics[width = .33\textwidth]{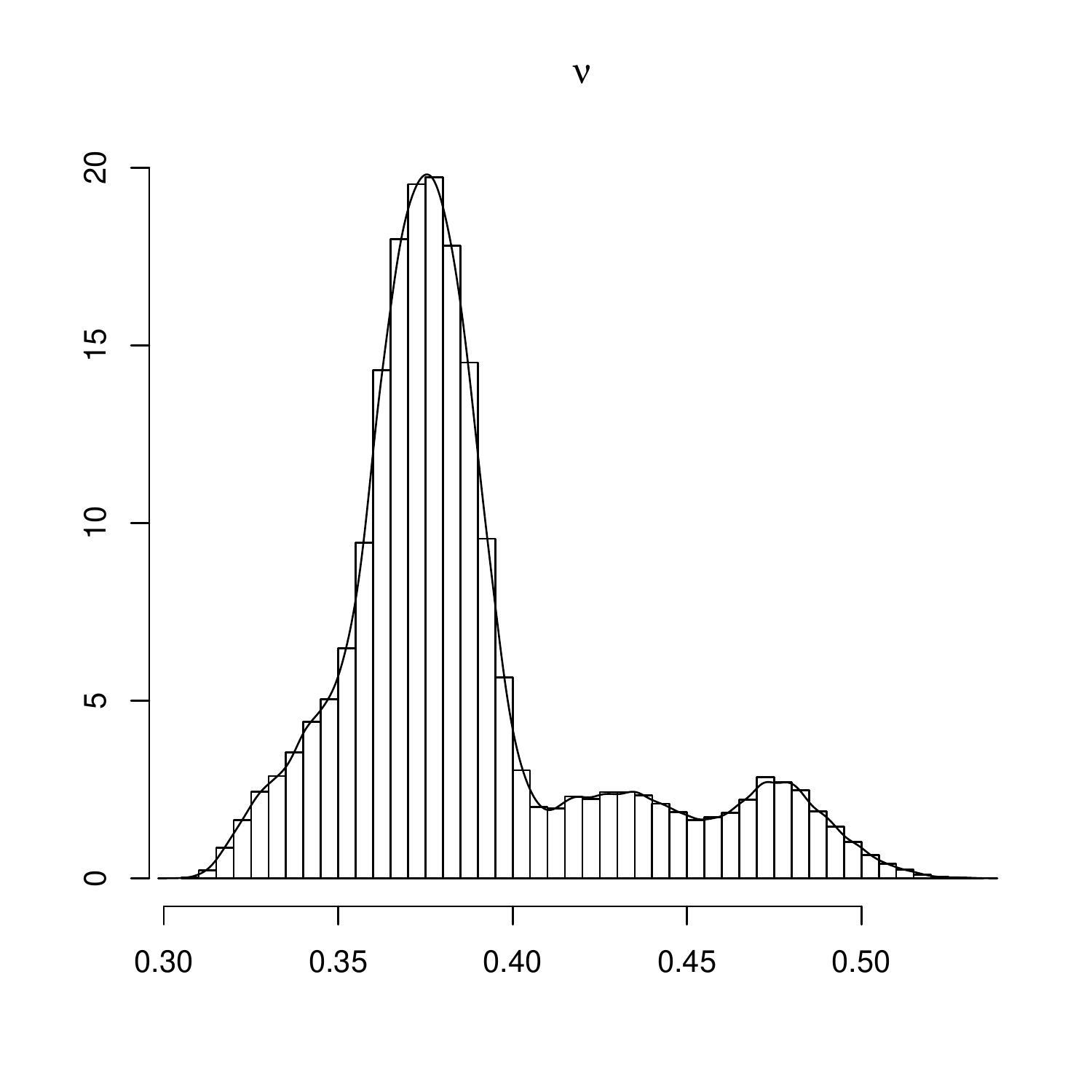} \\
(a) Ambulatory expenditures
&
(b) Wage offer
&
(c) HIV survey
\end{tabular}
\caption{Posterior Distribution of $\nu$ in Three Empirical Studies}\label{fg::nu}
\end{figure}

Due to the extremely heavy tails in the data ($\nu<1$ for most posterior draws in Figure \ref{fg::nu}(c)), the regression coeffficients under selection-Probit and selection-Robit models differ a lot. And the conclusion about selection effect is much more significant under the selection-Robit model than under the selection-Probit model. The presence of selection effect makes more practical sense in this particular example. Since the HIV positive individuals tend
to hide their status and reject to participate in the HIV test, the error terms $\varepsilon_i$ and $\eta_i$ should be negatively correlated.

\newsavebox{\tableboxboxbox}
\begin{lrbox}{\tableboxboxbox}
\begin{tabular}{| rrrrr |}
\hline
        & \multicolumn{2}{c}{ Selection-Probit (\texttt{heckprob}) } & \multicolumn{2}{c|}{ Selection-Robit (Bayes)} \\
        \hline
        \multicolumn{5}{|l|}{Outcome Model}\\
age &       $0.035$ & $(0.007, 0.065)$ &  $0.037$ & $(-0.016, 0.095)$\\
income&  $-0.157$ & $(-0.308, -0.007) $ & $-0.355$& $(-0.772, 0.070)$\\
marital &  $0.708$ & $(0.517, 0.898)$ &  $2.179$& $(1.407, 4.595)$\\
condom &$0.266$ & $(0.130, 0.402)$ & $0.715$& $(0.448, 1.011)$\\
evertestHIV  &$0.167$ & $(0.058, 0.277)$ & $0.297$ & $( 0.063, 0.534 )$ \\
smoking&$0.002$ & $(-0.113, 0.117) $& $-0.005$& $(-0.271, 0.248)$\\
\hline
        \multicolumn{5}{|l|}{Selection Model}\\
age &  $0.024$   & $(0.001, 0.046)$&  $0.303$&  $(0.045, 0.484)$\\
income& $-0.159$&$(-0.270, -0.047)$& $-0.762$& $(-1.489, 0.512)$\\
marital&$-0.074$ & $(-0.176, 0.029)$& $-1.001$& $(-1.941, 0.255)$\\
condom&$0.055$ & $(-0.041, 0.151)$& $-0.141$& $(-1.402, 0.672)$\\
evertestHIV&$0.045$ & $(-0.040, 0.130)$ & $0.540$& $(-0.135, 1.096)$\\
smoking &$0.160$&$(0.072, 0.248)$& $1.043$&  $(-0.268, 1.884)$\\
\hline
$\rho$  & $-0.590$ & $(-0.863, 0.125)$ &  $-0.968$ & $(-0.999, -0.737)$ \\
$\nu$ & $\infty$  & &  $0.378$ & $(0.330, 0.487)$\\
 \hline
\end{tabular}
\end{lrbox}

\begin{table}[t]
\caption{The HIV Survey Example}\label{tb::HIV}
\label{tb::HIV}
\begin{center}
\resizebox{0.6\textwidth}{!}{\usebox{\tableboxboxbox}}
\end{center}
{\small Note: MLEs and $95\%$ confidence intervals of the selection-Probit model (column 2 and 3); Bayesian posterior medians and $95\%$ credible intervals of the selection-Robit model (column 4 and 5).
}
\end{table}

\clearpage

%


\section{Discussion}\label{sec::discussion}

This paper develops new Bayesian procedures for a class of selection-t models, which was recently proposed by Marchenko and Genton (2012) to deal with heavy-tailedness of the data. Selection models with t errors are robust parametric alternatives for the selection models with Normal errors (Heckman 1979). Although selection-t models are not as flexible as semiparametric (Chib et al. 2009) and nonparametric (Van Hasselt 2011) models, they can model data with heavy tails by introducing only one extra parameter $\nu$ for controlling the heavy-tailedness (Marchenko and Genton 2012).
Efficient implementations of the selection models are realized by Bayesian procedures using data augmentation and parameter expansion. 
We illustrate the potential applications of our Bayesian procedures for selection-t models with three real problems.
The heavy-tailedness seems very common in practice, since
we find strong evidence of heavy-tailedness in all of our empirical studies.
In our examples, the conclusions about the existence of the selection effect differ dramatically under different distributional assumptions of the error terms, and results from different models may provide different practical interpretations.
It is our future research direction to study the generalized selection-t models with different degrees of freedom for the outcome equation and selection equation.

Due to the identification issue, the variances of the error terms in the selection equation in the selection and selection-t models are restricted to be one, and the variances of the error terms in both the selection and outcome equations in the selection-Probit and selection-Robit models are restricted to be one.
These restrictions make the posterior distributions of the covariance matrix nonstandard and nonconjugate, which complicate the MCMC procedures.
One possible solution of this problem is to reparametrize the covariance matrix (Koop and Poirier 1997; Li 1998; McCulloch, Polson and Rossi 2000; Van Hasselt 2011) in terms of the covariance cov$( \varepsilon_i, \eta_i) = \sigma_{12}$ and the conditional variance var$( \varepsilon_i\mid  \eta_i  ) = \sigma_{1|2}^2$: 
$$
\bm{\Omega} =  \begin{pmatrix}
\sigma_1^2 & \rho \sigma_1 \\
\rho \sigma_1 & 1
\end{pmatrix} = 
\begin{pmatrix}
\sigma_{1|2}^2 + \sigma_{12}^2 & \sigma_{12} \\
\sigma_{12} & 1
\end{pmatrix}.
$$
Imposing inverse-Gamma prior on $\sigma_{1|2}^2$ and Normal prior on $\sigma_{12}$ will result in conjugate conditional posterior distributions of the parameters. 
Our paper takes an alternative perspective, and the MCMC algorithms rely on the parameter expansion technique by enlarging the parameter space to maintain conjugacy, which ``is closely related to reparametrization techniques'' (Liu and Wu 1999). Thus, the previous reparametrization method and our parameter expansion method share some common essence. In analysis of the multinomial Probit model, 
Imai and Van Dyk (2005) used the parameter expansion method, and showed some advantages of the parameter expansion approach in terms of convergence rate of the MCMC algorithms.  
Although our algorithms require imputing all the missing $y_i^*$ which may adversely affect the mixing properties, the parameter expansion seems to be a compensation of this drawback in terms of convergence rate. 
Our future research will focus on combination the strength of both approaches. 
Although reparametrization in Koop and Poirier (1997), Li (1998), McCulloch, Polson and Rossi (2000) and Van Hasselt (2011) is very convenient to deal with selection models, the implied prior for $\rho$ is not as transparent as that implied by our Inverse-Wishart prior. As discussed before, $\bm{W}_2^{-1}(3, \bm{I}_2)$ is a marginally uniform prior (Barnard, McCulloch and Meng 2000), which implies that $\rho\sim $ Uniform$(-1,1)$.
Moreover, the reparametrization method is not directly applicable to the selection-Probit or selection-Robit, since the posterior distribution of $\rho$ is nonconjugate or has a nonstandard form, which, however, can be easily solved by
our parameter expansion scheme. The power of unification of all the selection models of parameter expansion motivates our Bayesian procedures.

In our first two empirical studies, frequentists' and Bayesian procedures give very similar results, which are consequences of the Bernstein-Von Mises theoreom (Van der Vaart, 2000) that the sampling distribution of the MLE and the posterior distribution for the same parameter have the same asymptotic Normality under regularity conditions and with large samples.
However, they may provide different results when the regularity conditions fail or with small samples.  
In our third empirical study, the posterior distribution of $\nu$ is not unimodal, and the posterior distribution of the parameter $\rho$ is very close to the boundary $-1$. The regularity conditions for asymptotic Normality may be violated, and inference based on Normal approximation may be inappropriate.

\section*{Acknowledgements}
I want to thank Dr. Yulia V. Marchenko for providing him with the \texttt{Stata} command \texttt{heckt}, which implements the procedure in Marchenko and Genton (2012). And I am also grateful to the reviewers' constructive comments.

\section*{Appendices}
\subsection*{Appendix A: Prior for $\bm{\Sigma}$ and Prior for $(\bm{\Omega}, \sigma_2^2)$}
\setcounter{equation}{0}
\renewcommand {\theequation} {A.\arabic{equation}}

Let $\omega_{ij}$ denote the $(i, j)-$th element of $\bm{\Omega}$ with $\omega_{11} = \sigma_1^2$, and $\omega_{12} = \rho\sigma_1$. Let $\sigma_{ij}$ denote the $(i, j)-$th element of 
$\bm{\Sigma}$ with $\sigma_{11} = \sigma^2_1 = \omega_{11}, \sigma_{12} = \rho\sigma_1\sigma_2 = \omega_{12} (\sigma_2^2)^{1/2}$ and $\sigma_{22} = \sigma^2_2$.
The Jacobian matrix of the transformation $ (\bm{\Omega}, \sigma_2^2)\rightarrow \bm{\Sigma}$ is
\begin{eqnarray*}
\frac{\partial (\sigma_{11}, \sigma_{12}, \sigma_{22})}{\partial (\omega_{11}, \omega_{12}, \sigma^2_2)}  = 
\bordermatrix{
                      &   \omega_{11}         &    \omega_{12}      &   \sigma_2^2       \cr
\sigma_{11}   &      1                         &  0                          & 0  \cr
\sigma_{12}   &      0                         &    \sigma_2           &    \omega_{12} /(2\sigma_2)\cr 
\sigma_{22}   &     0                          &      0                      & 1
},
\end{eqnarray*}
and therefore the Jacobian of the transformation is $J \{ ( \bm{\Omega}, \sigma_2^2 )\rightarrow \bm{\Sigma}\}= \sigma_2.$

If we use Inverse-Wishart prior $\bm{W}^{-1}_2(\nu_0, \bm{I}_2)$ for $\bm{\Sigma}$ with density
\begin{eqnarray*}
f(\bm{\Sigma}) \propto |\bm{\Sigma}|^{-(\nu_0 + 2 + 1)/2} \exp\left\{   -\frac{1}{2} \text{tr} \left(  \bm{\Sigma}^{-1}  \right) \right\},
\end{eqnarray*}
the prior for $(\sigma_2^2,  \bm{\Omega})$ is
\begin{eqnarray*}\label{eq::prior2}
f(\sigma_2^2,  \bm{\Omega})   \propto (  \sigma_2^2 )^{-\nu_0/2 - 1}  |\bm{\Sigma}|^{-(\nu_0 + 2+1)/2} \exp\left\{  -\frac{1}{2} \left( \omega^{11}  + \frac{\omega^{22}}{\sigma^2_2}    \right)   \right\},
\end{eqnarray*}
where $\omega^{ij}$ is the $(i, j)$-th element of $\bm{\Omega}^{-1}.$
From joint distribution of $(\bm{\Omega}, \sigma^2_2)$, 
the prior for $\sigma^2_2$ given $\bm{\Omega}$ is $\sigma^2_2|\bm{\Omega}\sim \omega^{22}/\chi^2_{\nu_0} = \{ (1- \rho^2)\chi^2_{\nu_0} \} ^{-1}$, and
the prior for $\bm{\Omega}$ is 
\begin{eqnarray*}
f(\bm{\Omega})   &\propto&    |\bm{\Omega} |^{-(\nu_0 + 3)/2}  e^{-\omega^{11}/2}  \left(\omega^{22}  \right)^{-\nu_0/2}\\
&\propto&   (1 - \rho^2)^{-3/2}  \sigma_1^{-(\nu_0 + 3)} \exp\left\{   -\frac{1}{ 2\sigma^2_1 (1 - \rho^2) }   \right\}.
\end{eqnarray*}


\subsection*{Appendix B: Prior for $\bm{\Sigma}$ and Prior for $(\sigma_1^2, \sigma_2^2, \bm{R})$}

The Jacobian matrix of the transfomation $(\sigma_1^2, \sigma_2^2, \bm{R})\rightarrow   \bm{\Sigma}$ is
\begin{eqnarray*}
\frac{\partial (\sigma_{11}, \sigma_{12}, \sigma_{22})}{\partial (\sigma_1^2, \sigma_2^2, \rho)}  = 
\bordermatrix{
                      &   \sigma_{1}^2         &    \rho      &   \sigma_2^2       \cr
\sigma_{11}   &      1                         &  0                          & 0  \cr
\sigma_{12} 
                    &      \rho (\sigma_2^2/\sigma_1^2)^{1/2}                         
                                                        &   \sigma_1 \sigma_2           
                                                                                        &   \rho(\sigma_1^2/\sigma_2^2)^{1/2 } \cr 
\sigma_{22}   &     0                          &      0                      & 1
},
\end{eqnarray*}
and therefore the Jacobian of the transformation is $J\{ (\sigma_1^2, \sigma_2^2, \bm{R})\rightarrow  \bm{\Sigma}  \} = \sigma_1\sigma_2.$

If we use Inverse-Wishart prior $\bm{W}^{-1}_2(\nu_0, \bm{I}_2)$ for $\bm{\Sigma}$, the prior for $(\sigma_1^2, \sigma_2^2, \bm{R})$ is
\begin{eqnarray*}
f(\sigma_1^2, \sigma_2^2, \bm{R}) \propto  (\sigma_1\sigma_2)^{-(\nu_0 + 1)} |\bm{R}|^{-(\nu_0 + 2+1)} 
\exp\left\{   - \frac{r^{11}}{2\sigma_1^2}  - \frac{r^{22}}{2\sigma_2^2}   \right\},
\end{eqnarray*}
where $r^{ij}$ is the $(i, j)$-th element of $\bm{R}^{-1}$.
From the joint distribution of $(\sigma_1^2, \sigma_2^2, \bm{R})$, the priors for $\sigma_1^2$ and $\sigma_2^2$ are $\sigma_1^2| \bm{R} \sim r^{11}/\chi^2_{\nu_0} = \{(1 - \rho^2 ) \chi^2_{\nu_0} \}^{-1}, \sigma_2^2|\bm{R} \sim r^{11}/\chi^2_{\nu_0} = \{(1 - \rho^2 ) \chi^2_{\nu_0} \}^{-1}$, and they are independent given $\bm{R}$.
The prior for $\bm{R}$ or equivalently $\rho$ is
$f(\rho) \propto ( 1 - \rho^2  )^{- (\nu_0 - 3)/2}.
$

\subsection*{Appendix C: Gamma Approximation}
Ignoring additive constants, the log conditional density of $\nu$ is
\begin{eqnarray*}
l(\nu) =  N \nu \log(\nu/2) /2  - N \log \Gamma(\nu/2)  + (\alpha_0 - 1)\log \nu - \xi \nu   ,
\end{eqnarray*}
and the log density of Gamma$(\alpha^*, \beta^*)$ is
\begin{eqnarray*}
h(\nu) =   (\alpha^* - 1)\log \nu - \beta^* \nu.
\end{eqnarray*}

The first and second order derivatives of $l(\nu)$ and $h(\nu)$ are
\begin{eqnarray}
&l'(\nu) = \frac{N}{2}  \log \left( \frac{\nu}{2} \right) + \frac{N}{2} - \frac{N}{2} \psi \left( \frac{ \nu}{2} \right) 
+ \frac{\alpha_0 - 1}{\nu} - \xi,
&h'(\nu) = \frac{\alpha^* - 1}{\nu} - \beta^*, \label{eq::foc}\\
&l''(\nu) = \frac{N}{2\nu} - \frac{N}{4} \psi'\left(\frac{\nu}{2}\right) - \frac{\alpha_0 - 1}{\nu^2},
&h''(\nu) = -\frac{\alpha^* - 1}{\nu^2},\label{eq::soc}
\end{eqnarray}
where $\psi(x) = \d \log\Gamma(x)/\d x $ is the digamma function, and $\psi'(x) = \d\psi(x)/\d x$ is the trigamma function.
The mode of $h(\nu)$ is $  (\alpha^* - 1)/\beta^* $, and the curvature at the mode is $- \beta^{*2}/(\alpha^* - 1).$
From (\ref{eq::foc}) and (\ref{eq::soc}), we can numerically find the mode of $l(\nu)$ denoted as $\nu^*$, and the curvature at the mode is $l^* = l''(\nu^*)$. 
By matching the modes and the curvatures at the modes of $l(\nu)$ and $h(\nu)$, the parameters of the Gamma approximation is chosen as 
\begin{eqnarray}\label{eq::nu}
\alpha^* = 1 - \nu^{*2}l^*  \text{ and } \beta^* =  - \nu^*l^*.
\end{eqnarray}

\section*{References}
\begin{description} \itemsep=-\parsep \itemindent=-1.2 cm

\item Ahn, H. and Powell, J. L. (1993), ``Semiparametric Estimation of Censored Selection Models with a Nonparametric Selection Mechanism,'' {\it Journal of Econometrics}, {\bfseries 58}, 3-29.



\item Barnard, J., McCulloch, R. and Meng, X. L. (2000), ``Modeling covariance matrices in terms of standard deviations and correlations, with application to shrinkage,'' {\it Statistica Sinica}, {\bfseries 10}, 1281-1311.

\item B\"arnighausen T., Bor J., Wandira-Kazibwe S., Canning D. (2011), ``Correcting HIV prevalence estimates for survey nonparticipation using Heckman-type selection models,''  {\it Epidemiology}, {\bfseries 22}, 27-35.


\item Cameron, A. C. and Trivedi, P. K. (2010), {\it Microeconometrics Using Stata} (Revised Edition), College Station, TX: Stata Press.


\item Chib, S., Greenberg, E. and Jeliazkov, I. (2009), ``Estimation of Semiparametric Models in the Presence of Endogeneity and Sample Selection,'' {\it Journal of Computational and Graphical Statistics}, {\bfseries 18}, 321-348.


\item Das, M., Newey, W. K. and Vella, F. (2003), ``Nonparametric Estimation of Sample Selection Models,'' {\it Review of Economics Studies}, {\bfseries 70}, 33-58.


\item Geweke, J. (1992), ``Priors for Macroeconomic Time Series and Their Application,'' Minneapolis: Federal Reserve Bank of Minneapolis, Institute for Empirical Macroeconomics discussion paper \#64.


\item Heckman, J. J. (1979), ``Sample Selection Bias as a Specification Error,'' {\it Econometrica}, {\bf 47}, 153-161.

\item Imai, K. and Van Dyk, D. A. (2005), ``A Bayesian Analysis of the Multinomial Probit Model Using Marginal Data Augmentation,'' {\it Journal of Econometrics}, {\bfseries 124}, 311-334.



\item Lee, L. F. (1983), ``Generalized Econometric Models with Selectivity,'' {\it Econometrica}, {\bfseries 51}, 507-512.


\item Li, K. (1998), ``Bayesian Inference in a Simultaneous Equation Model with Limited Dependent Variables,'' {\it Journal of Econometrics}, {\bfseries 85}, 387-400.

\item Little, R. J. and Rubin, D. B. (2002), {\it Statistical Analysis of Missing Data}, Wiley: New York.

\item Liu, C. (2004), ``Robit Regression: A Simple Robust Alternative to Logistic and Probit Regression,'' In {\it Applied Bayesian Modeling and Causal Inference from Incomplete-Data Perspectives} (A. Gelman and X. L. Meng, eds.) 227-238. Wiley, New York.

\item Liu, C. (1999), ``Efficient ML Estimation of the Multivariate Normal Distribution from Incomplete Data,'' {\it Journal of Multivariate Analysis}, {\bfseries 69}, 206-217.

\item Liu, C., Rubin, D. B. and Wu, Y. N. (1998), ``Parameter Expansion to Accelerate EM: the PX-EM algorithm,'' {\it Biometrika}, {\bfseries 85}, 755-770.

\item Liu, J. S. (2001), {\it Monte Carlo Strategies in Scientific Computing}, NY: Springer-Verlag.

\item Liu, J. S. and Wu, Y. N. (1999), ``Parameter Expansion for Data Augmentation,'' {\it Journal of the American Statistical Association}, {\bfseries 94}, 1264-1274.

\item Koop, G. and Poirier, D. J. (1997) Learning about the across-regime correlation in switching regression models. {\it Journal of Econometrics}, {\bfseries 78}, 217-227.


\item Marchenko, Y. V. and Genton, M. G. (2012), ``A Heckman Selection-t Model,'' {\it Journal of the American Statistical Association}, {\bfseries 107}, 304-317.

\item McCulloch, R. E., Polson, N. G. and Rossi, P. E. (2000), ``A Bayesian Analysis of the Multinomial Probit Model with Fully Identified Parameters,'' {\it Journal of Econometrics}, {\bfseries 99}, 173-193.

\item Meng, X. L. and Van Dyk, D. A. (1999), ``Seeking Efficient Data Augmentation Schemes via Conditional and Marginal Augmentation,'' {\it Biometrika}, {\bfseries 86}, 301-320.

\item Mroz, T. A. (1987), ``The Sensitivity of an Empirical Model of Married Women’s Hours of Work to Economical and Statistical Assumptions,'' {\it Econometrica}, {\bfseries 55}, 765-799.

\item
R Development Core Team (2010). R: A language and environment for statistical computing. R Foundation
for Statistical Computing, Vienna, Austria. ISBN 3-900051-07-0, URL \texttt{http://www.R-project.org/}.

\item
StataCorp. (2013). Stata Statistical Software: Release 13. College Station, TX: StataCorp LP.





\item Tanner, M. A., and Wong, W. H. (1987), ``The Calculation of Posterior Distributions by Data Augmentation'' (with discussion), {\it Journal of the American Statistical Association}, {\bfseries 82}, 528-550. 
  
\item Toomet, O. and Henningsen, A. (2008), ``Sample Selection Models in \texttt{R}: Package \texttt{sampleSelection},'' {\it Journal of Statistical Software}, {\bfseries 27}, \url{http://www.jstatsoft.org/v27/i07}.

\item
Van der Vaart, A. W. (2000). {\it Asymptotic Statistics.} Cambridge University Press.

\item Van Dyk, D. A. and Meng, X. L. (2001), ``The Art of Data Augmentation,'' {\it Journal of Computational and Graphical Statistics}, {\bfseries 10}, 1-50.

\item Van Hasselt, M. (2011), ``Bayesian Inference in a Sample Selection Model,'' {\it Journal of Econometrics}, {\bfseries 165}, 221-232.



\item Wooldridge, J. M. (2002), {\it Econometric Analysis of Cross Section and Panel Data}, The MIT press.

\end{description}

\end{document}